\documentclass[aps,preprint,floatfix,nofootinbib,showpacs]{revtex4-1}
\pdfoutput=1
\usepackage{dcolumn}
\usepackage{bm}
\usepackage{graphicx}
\usepackage{amssymb,amsmath}
\usepackage{multirow}
\usepackage{units,changes}
\usepackage{color,url}
\usepackage{tabu}
\usepackage{array}
\usepackage[colorlinks=true,urlcolor=blue,anchorcolor=blue
,citecolor=blue,filecolor=blue,linkcolor=blue,menucolor=blue
,linktocpage=true,pdfproducer=medialab,pdfa=true]{hyperref}
\usepackage{colordvi}
\usepackage{subcaption}
\def\beqa{\begin{eqnarray}}
\def\eeqa{\end{eqnarray}}

\usepackage{soul}
\usepackage{hhline} 
\newcommand{\apjs}{Astrophys. J. Suppl.}

\newcommand{\gev}{\ensuremath{\,\mathrm{GeV}}}

\newcommand{\pc}{\ensuremath{\,\mathrm{pc}}}
\newcommand{\kpc}{\ensuremath{\,\mathrm{kpc}}}

\newcommand{\sv}{\ensuremath{\langle\sigma v_{\rm rel.}\rangle}}
\newcommand{\dnde}{\ensuremath{\frac{dN_\gamma}{dE_\gamma}}} 
\newcommand{\avgdnde}{\ensuremath{\langle dN_\gamma/dE_\gamma\rangle[r]}}

\begin{document}
\title{Exploring semi-relativistic $p$-wave dark matter annihilation in minimal Higgs portal near supermassive black hole}
\def\slash#1{#1\!\!\!/}
\author{Chih-Ting Lu$^1$~\footnote{Electronic address: ctlu@njnu.edu.cn}, Xiao-Yi Luo$^1$~\footnote{Electronic address: xyluo@nnu.edu.cn}, Zi-Qing Xia$^{2}$~\footnote{Electronic address: xiazq@pmo.ac.cn}}
\affiliation{$^1$Department of Physics and Institute of Theoretical Physics, Nanjing Normal University, Nanjing, 210023, China} 
\affiliation{$^2$Key Laboratory of Dark Matter and Space Astronomy, Purple Mountain Observatory, Chinese Academy of Sciences, Nanjing 210033, China}

\date{\today}

\begin{abstract}

We conduct a comprehensive analysis of potential annihilation processes of light dark matter (DM) in minimal Higgs portal models near supermassive black hole (Sgr A$^{\star}$) in the Galactic Center, considering interactions between DM particles mediated by either a light scalar or pseudoscalar with couplings $ c_s $ and $ c_p $. Accelerated by the supermassive black hole, DM particles can reach velocities up to half the speed of light, significantly enhancing the $ p $-wave annihilation cross-section, allowing forbidden annihilation channels within specific mass ranges, and producing unique gamma-ray spectral signals. 
Utilizing gamma-ray observation from Fermi Large Area Telescope (Fermi-LAT) in the direction of Sgr $A^{\star}$, we constrain light DM parameter in the mass range of $ 0.3-10 \, \text{GeV} $ . Our results indicate that the couplings $ c_s $ and $ c_p $ are constrained to the order of $ 10^{-5} $, corresponding to a DM annihilation cross-section as low as $ 10^{-38} $$ {\rm cm}^3/{\rm s}$. 
In the future, the Very Large Gamma-ray Space Telescope (VLAST), with a larger detection area and broader detection range from $1$ MeV to $1$ TeV, will enhance our ability to probe sub-GeV DM and offer the opportunity to further study the forbidden annihilation scenario.

\end{abstract}

\maketitle
\section{Introduction} 

Dark matter (DM), an elusive and non-luminous component of the universe, was first suggested to explain discrepancies in astronomical observations~\cite{Zwicky:1933gu,Zwicky:1937zza,Rubin:1970zza,Massey:2010hh}. The Cosmic Microwave Background (CMB) radiation, the afterglow of the Big Bang, also supports the existence of DM through its influence on the temperature fluctuations in the early universe~\cite{Planck:2018vyg}. DM is essential to our understanding of the universe because it explains several gravitational phenomena that cannot be accounted for by ordinary (baryonic) matter alone. Additionally, DM is crucial for explaining the anisotropies in the CMB and the overall shape and expansion rate of the universe, as described by the Lambda Cold Dark Matter ($\Lambda$CDM) model~\cite{Condon:2018eqx}.

However, the exact nature of DM remains one of the biggest mysteries in cosmology and particle physics. It is hypothesized to be composed of non-baryonic particles that do not interact with electromagnetic forces, making them invisible and difficult to detect. Several DM candidates have been proposed, including Weakly Interacting Massive Particles (WIMPs)~\cite{Arcadi:2017kky,Arcadi:2024ukq}, axions~\cite{Duffy:2009ig,Chadha-Day:2021szb}, and sterile neutrinos~\cite{Boyarsky:2009ix,Boyarsky:2018tvu}. Among these, WIMPs are a popular candidate because they naturally arise in some well-motivated models and can account for the correct relic abundance of DM from the early universe~\cite{Jungman:1995df,Birkedal:2006fz,Hooper:2007qk}. For WIMPs, the mass range is typically thought to be between a few GeV to a few TeV, with interaction strengths weak enough to avoid detection in ordinary matter but sufficient to leave a detectable signature in DM direct detection~\cite{Schumann:2019eaa}, DM indirect detection~\cite{Gaskins:2016cha}, and collider~\cite{Kahlhoefer:2017dnp,Boveia:2018yeb} experiments. Although WIMPs are predictive and detectable, we have not yet found any concrete evidence of them. Therefore, researchers are extending their attention to phenomenological models and detection strategies for lighter (sub-GeV) DM~\cite{Knapen:2017xzo,Lin:2019uvt}.

According to the Lee-Weinberg bound~\cite{Lee:1977ua}, sub-GeV DM via thermal production cannot satisfy the relic abundance without including a new light mediator. Therefore, predictions of light mediators, such as dark photons~\cite{Fabbrichesi:2020wbt}, dark scalars~\cite{Clarke:2013aya}, and axion-like particles~\cite{Choi:2020rgn}, become a common feature in many sub-GeV DM models~\cite{Krnjaic:2015mbs,Darme:2017glc,Bharucha:2022lty,Balan:2024cmq}. Additionally, the sub-GeV DM annihilation cross section faces stringent constraints from the CMB~\cite{Planck:2018vyg,Slatyer:2009yq,Sabti:2019mhn}, prompting researchers to explore alternative mechanisms such as $p$-wave DM annihilation~\cite{Boehm:2003hm,Kumar:2013iva}, resonant DM annihilation~\cite{Griest:1990kh,Binder:2022pmf}, and forbidden DM annihilation~\cite{Griest:1990kh,DAgnolo:2015ujb} and others~\cite{Lin:2011gj,Hochberg:2014dra,Izaguirre:2015zva,Kuflik:2015isi,Cline:2017tka,DAgnolo:2017dbv,Frumkin:2021zng}. The minimal Higgs portal model~\cite{Arcadi:2019lka,Matsumoto:2018acr,Bondarenko:2019vrb,Chen:2024njd} is one of the simplest sub-GeV DM models, including a new fermionic field as the DM candidate and a new scalar singlet field as the mediator. Some proposals to explore this model include X-ray emission and gamma-ray telescope searches~\cite{Cirelli:2020bpc,Cirelli:2023tnx,Chen:2024njd}, cosmic-ray upscattering searches~\cite{Bondarenko:2019vrb,Bell:2023sdq}, and collider and fixed target searches~\cite{Batell:2019nwo,Filimonova:2019tuy,Batell:2020vqn,Ema:2020ulo}. However, all these detections require large enough couplings between DM and Standard Model (SM) particles.

In this work, we focus on the nightmare scenario in the minimal Higgs portal model, where only feeble interactions between DM and SM particles are considered. In this scenario, the DM is often referred to as secluded DM~\cite{Pospelov:2007mp,Pospelov:2008zw}. The observed DM relic abundance can be achieved if a pair of DM particles annihilate into a pair of mediators through thermal freeze-out. In this situation, only DM indirect detections are possible, but it remains challenging to detect sub-GeV DM annihilation with current observations. Moreover, this DM annihilation process suffers from 
$p$-wave suppression. Therefore, a novel strategy to detect such elusive light DM signals is urgently needed.


The vicinity of the supermassive black hole (SMBH) at the Galactic Center (GC), named as Sgr A$^{\star}$, is an optimal region for detecting velocity-dependent DM annihilation signalse~\cite{Johnson:2019hsm,Shelton:2015aqa}. The enormous gravitational potential of SMBHs is enhanced by their accretion processes, leading to a significant accumulation of DM particles and resulting in a pronounced density spike. Typically, it is assumed that the DM halo of the Milky Way is spherically symmetric and that there is no relative rotational motion between the DM halo and the galactic disk. For the DM velocity distribution, an isotropic Maxwell-Boltzmann distribution is commonly used, known as the Standard Halo Model. In this model, the DM average velocity in the solar system is approximately $ 220 \, \text{km/s} $, or about $ 10^{-3} c $~\cite{Kuhlen:2012ft}. However, in the gravitational field of SMBH, DM particles near the GC exhibit substantially higher velocities. As they approach the black hole, their maximum velocity can reach half-relativistic speeds (approximately $ v \approx 0.5 c $). Given that the $ p $-wave DM annihilation cross-section is proportional to $ v^2 $, this increase in velocity significantly enhances the DM annihilation cross-section and the potential for detection. Moreover, once the total kinetic energy exceeds the mass of the final state particles, this velocity enhancement can also open previously forbidden annihilation channels. Therefore, the environment surrounding the SMBH provides an ideal setting for investigating sub-GeV secluded DM $ p $-wave annihilation and forbidden annihilation processes.

The Large Area Telescope aboard the Fermi satellite (Fermi-LAT)~\cite{Fermi-LAT:2009ihh} is one of the most sensitive high-energy gamma-ray telescopes currently in operation, detecting gamma rays with energies spanning from below $100$ MeV to above $1$ TeV.
Since the launch of the Fermi satellite in 2008, many efforts have been dedicated to the DM indirect detection through its gamma-ray observations~\cite{Fermi-LAT:2013thd,Zhou:2014lva,Calore:2014xka,Fermi-LAT:2015att,Liang:2016pvm,Fermi-LAT:2016afa,Zhu:2022tpr,Yang:2024jtp}.
Operating in all-sky survey mode, Fermi-LAT has collected a vast amount of observation data towards the Galactic Center, which allows us to look for potential sub-GeV dark matter signals using it's gamma-ray observation from Sgr A$^{\star}$.

This study revisits the process of enhancement of concealed DM annihilation due to accretion near SMBHs, with a focus on two scenarios within the minimal Higgs portal model, specifically involving scalar $ h_s $ and pseudo-scalar $ h_p $ as mediators. We analyze the spectra and annihilation cross-sections from $ p $-wave annihilation and forbidden annihilation of sub-GeV DM particles near the SMBH, considering both position and velocity dependent effects. Utilizing $16$ years of Fermi-LAT data from Sgr A$^{\star}$, we search for spectral signals to place constraints on the coupling constants $ c_s $ and $ c_p $ between DM particles with scalar and pseudoscalar, respectively. 

The remainder of this article is organized as follows: Sec.~\ref{sec:model} provides a concise overview of the minimal Higgs portal DM model, emphasizing the characteristics of light mediators and their interactions with DM. Sec.~\ref{sec:halo} discusses the DM density profile and outlines the velocity-averaged annihilation cross-sections and spectra used to compute DM annihilation mechanisms. In Sec.~\ref{sec:GC}, we perform precise calculations of the gamma-ray flux near Sgr A$^{\star}$, accounting for both the velocity and positional dependencies of the DM annihilation cross-section and photon yield spectrum. Our numerical methodology and results are presented in Sec.~\ref{sec:data}. Finally, Sec.~\ref{sec:sum} summarizes and discusses our findings.

\section{The minimal Higgs portal dark matter model} 
\label{sec:model}

In this study, we consider a Dirac fermion dark matter (DM) particle $ \chi $ that interacts with the Standard Model (SM) sector via a real singlet scalar $ h_s $ or pseudoscalar $ h_p $ ,where $ c_p $ and $ c_s $ denote the respective coupling constants of their interactions with a pair of DM particles. The renormalizable Lagrangian for this minimal Higgs portal DM model can be expressed as 
\begin{eqnarray}
\label{eq:DM-Phi}
    {\cal L} = {\cal L}_{SM}+
    \bar{\chi}(i{\:/\!\!\!\! {\partial}}-m_{\chi})\chi
    +\frac{1}{2}(\partial{\Phi})^2-\frac{c_{s}}{2}\Phi\bar{\chi}\chi
    -\frac{c_{p}}{2}\Phi\bar{\chi}i\gamma_{5}\chi-V(\Phi,H)
\end{eqnarray}
where $ \mathcal{L}_{SM} $ represents the Lagrangian of the SM, and $ m_\chi $ denotes the mass of the DM particle. The scalar field $ \Phi $, along with the SM Higgs field $ H $, contributes to the scalar potential $ V(\Phi, H) $ of the model.  After spontaneous symmetry breaking, the fields $ H $ and $ \Phi $ can be expressed as $H=[0,(v_H+h')/\sqrt{2}]^T$ and $\Phi=v_\Phi+\phi'$, where $ v_H $ and $ v_\Phi $ represent the vacuum expectation values of the fields $ H $ and $ \Phi $, respectively. For simplicity, we can set $ v_\Phi $ to zero without loss of generality~\cite{Matsumoto:2018acr}. 
The physical eigenstates $(h', \phi')$ are transformed into mass eigenstates, where the scalar case corresponds to $(h, h_s)$ and the pseudoscalar case corresponds to $(h, h_p)$. This diagonalization process introduces a mixing angle $ \theta $ between these two physical eigenstates which depends on the parameters in the potential $ V(\Phi,H) $ and $ v_H $~\cite{Matsumoto:2018acr,Chen:2024njd}. 
 Therefore, we take $ m_{\chi}$, $c_p$, $c_s$, $m_{h_s}$, $m_{h_p}$,  and $ \sin \theta $ as input model parameters. The interactions between the scalar sector $(h, h_{s,p})$ and the DM sector can be expressed as
\begin{equation}
	\mathcal{L}_{\rm int} \supset
	- \frac{\cos\theta}{2}(c_s\,h_s\,\bar{\chi}\chi+c_p\,h_p\,\bar{\chi} i \gamma_5 \chi)
	+ \frac{\sin\theta}{2}(c_s\,h\,\bar{\chi}\chi+c_p\,h\,\bar{\chi} i \gamma_5 \chi).
\end{equation} 
According to the above expression, it is evident that the interaction between $ h $ and $ \chi $ is suppressed by $ \sin \theta $. Here we set $\sin \theta = 10^{-5}$ as a benchmark value for simplicity~\cite{Chen:2024njd}. 
Additionally, since $ \cos \theta $ is approximately equal to 1, the interaction strength of the mediator $ h_{s,p} $ with a pair of DM particles depends solely on the coupling constants $ c_s $ and $ c_p $. 

\begin{figure}[tb]
\centering{\includegraphics[width=0.6\textwidth]{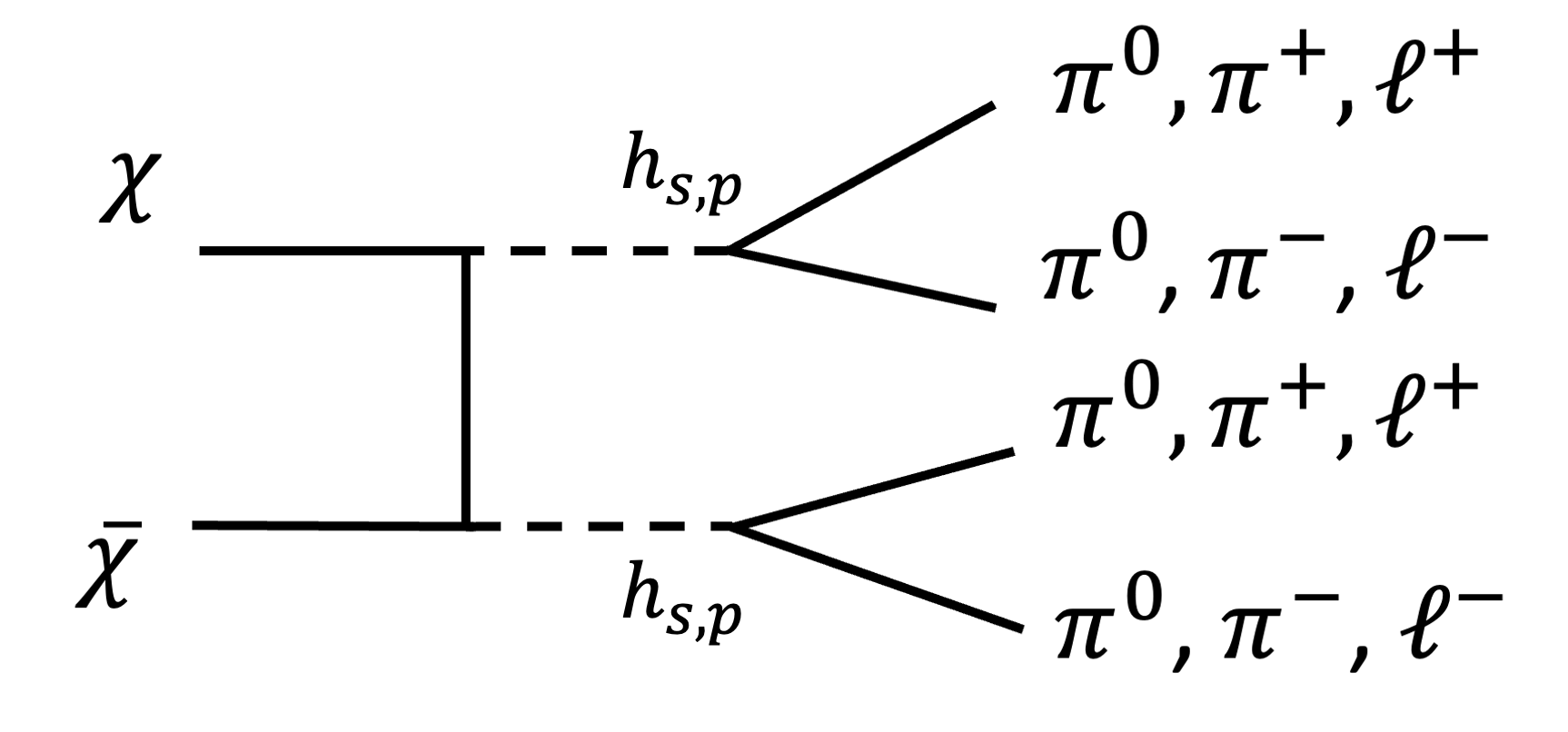}}
\caption{The DM annihilation processes to a pair of scalar mediators $h_s$ or  pseduoscalar mediators $h_p$. These two mediators subsequently decay into SM particles such as $\pi^0$, $\pi^{\pm}$ and $\ell^{\pm}$. Photons may be produced either through final state radiation from the charged leptons, $\pi^{\pm}$ or via the decay of $\pi^0$. 
}
\label{fig:anni}
\end{figure}

In the present model, we specifically focus on the secluded DM scenario~\cite{Pospelov:2007mp,Pospelov:2008zw} where the annihilation process $\chi \overline{\chi} \rightarrow h_{s,p}h_{s,p}$ occurs, and the annihilation cross-section exhibits dependence on the relative DM velocity ($p$-wave). Moreover, the velocity dependence of the DM annihilation cross-section can be significantly influenced by varying the mass ratios of the $\chi$ and $h_{s,p}$ particles. If the DM particles initially annihilate into a pair of mediators, which subsequently annihilate into SM particles such as light mesons ($\pi^0$, $\pi^{\pm}$), and charged leptons $\ell^{\pm}$ as illustrated in Fig.~\ref{fig:anni}, photons may be produced either through final state radiation (FSR) from the charged leptons as well as $\pi^{\pm}$ or via decay of $\pi^0$. 
There are two major Feynman diagrams of the DM annihilation processes, $\chi\bar{\chi}\to h_{s,p}  h_{s,p}$, including both $t$-channel and $u$-channel with $h_{s,p}$ exchange~\footnote{The three-point interactions involving $ h_s $ and $ h_p $ are assumed to be negligible for simplicity; therefore, contributions from the $s$-channel are not considered in this study.}. They can naturally relate to the forbidden DM scenarios ($m_\chi\lesssim m_{hs}$)~\cite{Griest:1990kh,DAgnolo:2015ujb,Delgado:2016umt,DAgnolo:2020mpt,Hara:2021lrj,Wojcik:2021xki}  or secluded DM scenarios ($m_\chi> m_{hs}$)~\cite{Pospelov:2007mp,Diamanti:2013bia,Bondarenko:2019vrb}. Qualitatively speaking, the annihilation cross-section of this channel exhibits the dominant contribution of $p$-wave in non-relativistic expansion, where the cross-section is proportional to the square of the relative velocity of DM particles.

We primarily investigated light DM with masses in the range of $ 0.3 - 10 \, \text{GeV} $. 
In the $p$-wave annihilation scenario, we set the DM mass larger than the mediator mass, $ m_{\chi} > m_{h_s,h_p} $. In the forbidden annihilation scenario, we considered two mass relations: $ m_{\chi} = 0.95 m_{h_s,h_p} $ and $ m_{\chi} = 0.85 m_{h_s,h_p} $. 
We then scanned the parameter space for coupling constants, $c_{s,p}$, in our analysis.
Previous studies have provided the allowed parameter space for the aforementioned parameters at a $95\%$ confidence level through global fitting~\cite{Matsumoto:2018acr,Chen:2024njd}. Specifically, it was found that for the $p$-wave annihilation scenario, the possible mediator mass range is from $ 5 \, \text{MeV} $ to $ 10 \, \text{GeV} $; for the forbidden annihilation scenario, the possible mediator mass range is from $ 0.32 - 11.76 \, \text{GeV} $. Within the DM mass range we consider, the coupling constant $ c_s $ for DM produced through the thermal freeze-out mechanism falls within the range of $ 3 \times 10^{-3} $ to $ 1 $. Additionally, in the above mediator mass range, the allowed range for $ \sin \theta $ is $ 10^{-10} $ to $ 0.1 $, making our choice of $ \sin \theta = 10^{-5} $ reasonable. 

Additionally, we calculate the DM self-interaction cross-section and find that, within the ranges of model parameters considered in this study, $\sigma_{\text{self}}/m_{\chi}$ reaches a maximum of $10^{-3}\ \text{cm}^2/\text{g}$. Since this value is so small, the effects of a strong self-interacting DM halo can be safely neglected in this work~\cite{Shapiro:2014oha}. The detailed calculation process is provided in Appendix~\ref{sec:self}.

\section{DM annihilation near the SMBH}
\label{sec:halo}



Considering the vicinity of the SMBH, the gravitational potential of the black hole dominates. As the distance from the black hole increases, its gravitational potential becomes weaker, and the gravitational potential of DM takes over when $r > 27$ pc. Using the method outlined in Ref.~\cite{Yang:2024jtp}, we calculate the potential energy function $\Phi(r)$ at the black hole's center, and then derive the phase space distribution of the DM based on this potential. Finally, we obtain the complete DM density profile near the galactic center by 
integrating the phase space function.


The obtained DM density profile is divided into four main segments: (1) Within the capture region (twice the Schwarzschild radius, $2R_{\rm s}$), the density of DM particles is zero; (2) Within the annihilation radius, $r_{\rm ann}$, the density decreases due to collisions between DM particles, resulting in an annihilation platform (cusp); (3) Outside the annihilation radius, the black hole's gravitational potential dominates, and the density profile forms a spike; (4) Beyond the potential energy boundary radius, the DM's gravitational potential dominates, and the density distribution follows the Navarro-Frenk-White (NFW) profile, given by $\rho_{\rm NFW}(r)=\rho_s(r/r_s)^{-\gamma}(1+r/r_s)^{\gamma-3}$ where $\rho_s\approx 0.0156 \; \rm M_{\odot}/\rm pc^3$, $r_s=1.5\times 10^4 \; \rm pc$, and $\gamma=0.8$, based on the S-star combined constraints $\gamma<0.83$~\cite{Shen:2023kkm}. We summarize the DM density profile from the above four segments as follows: 
\begin{equation}
    \rho(r)=\left\{
    \begin{aligned}
    &0, \quad\quad\quad\quad r<2R_{\rm s}, (\text{Capture Region}),\\
    &\rho_{\rm ann}(r), \quad  2R_{\rm s}<r<r_{\rm ann},(\text{cusp}),\\
    &\rho_{\rm spike}(r), \quad r_{\rm ann}<r<r_{\rm sp},(\text{spike}),\\
    &\rho_{\rm NFW}(r), \quad r>r_{\rm sp},(\text{NFW}),
    \end{aligned}
    \right 
.
\end{equation}
where $R_s$ is the Schwarzschild radius, $r_{\rm ann}$ is the annihilation radius, and $r_{\rm sp}=27$ pc is the the boundary radius where the black hole's potential energy dominates, as calculated in Ref.~\cite{Yang:2024jtp}.

\begin{figure*}[ht]
\centering
\includegraphics[width=7.8cm]{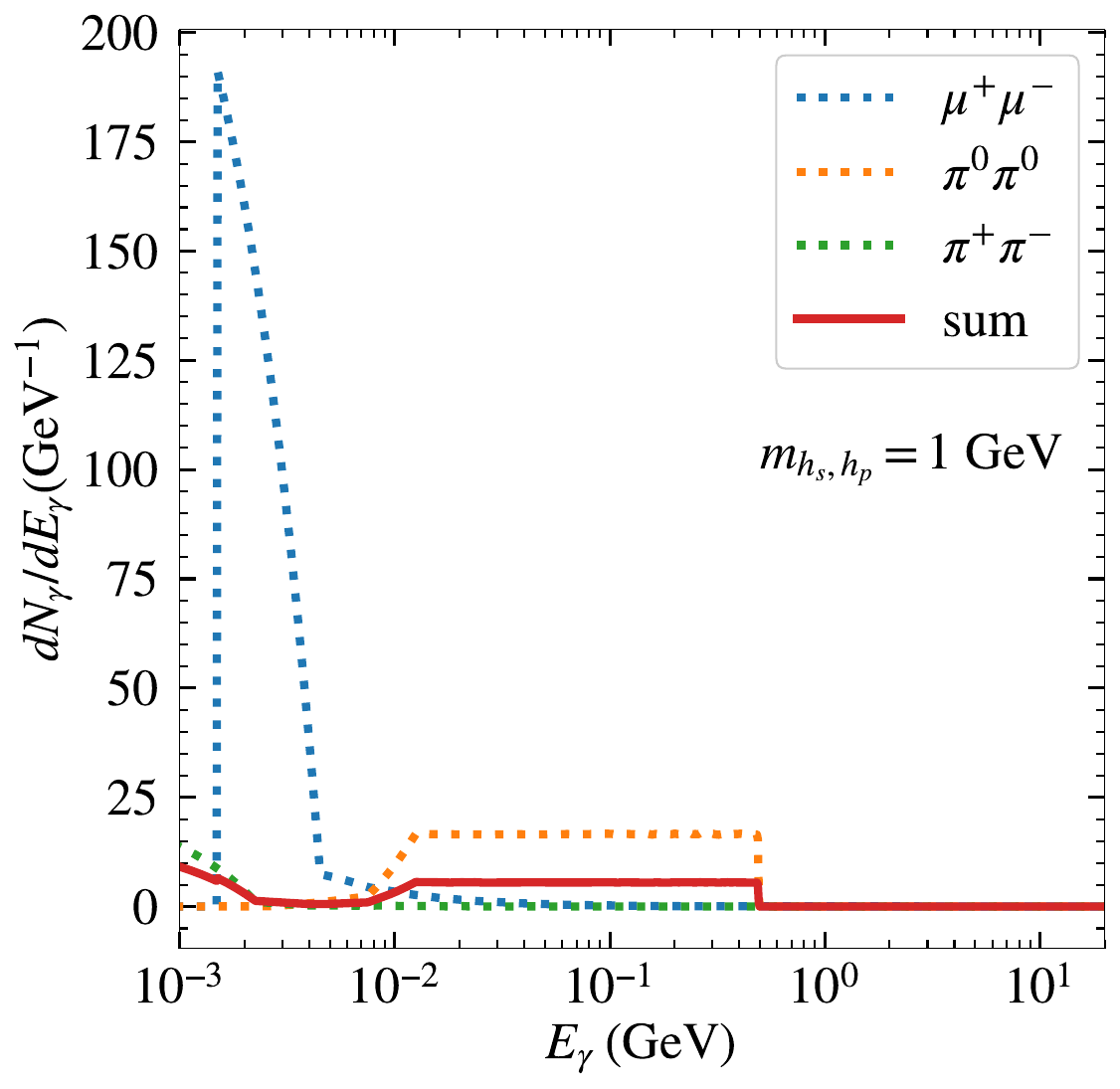}
\includegraphics[width=7.8cm]{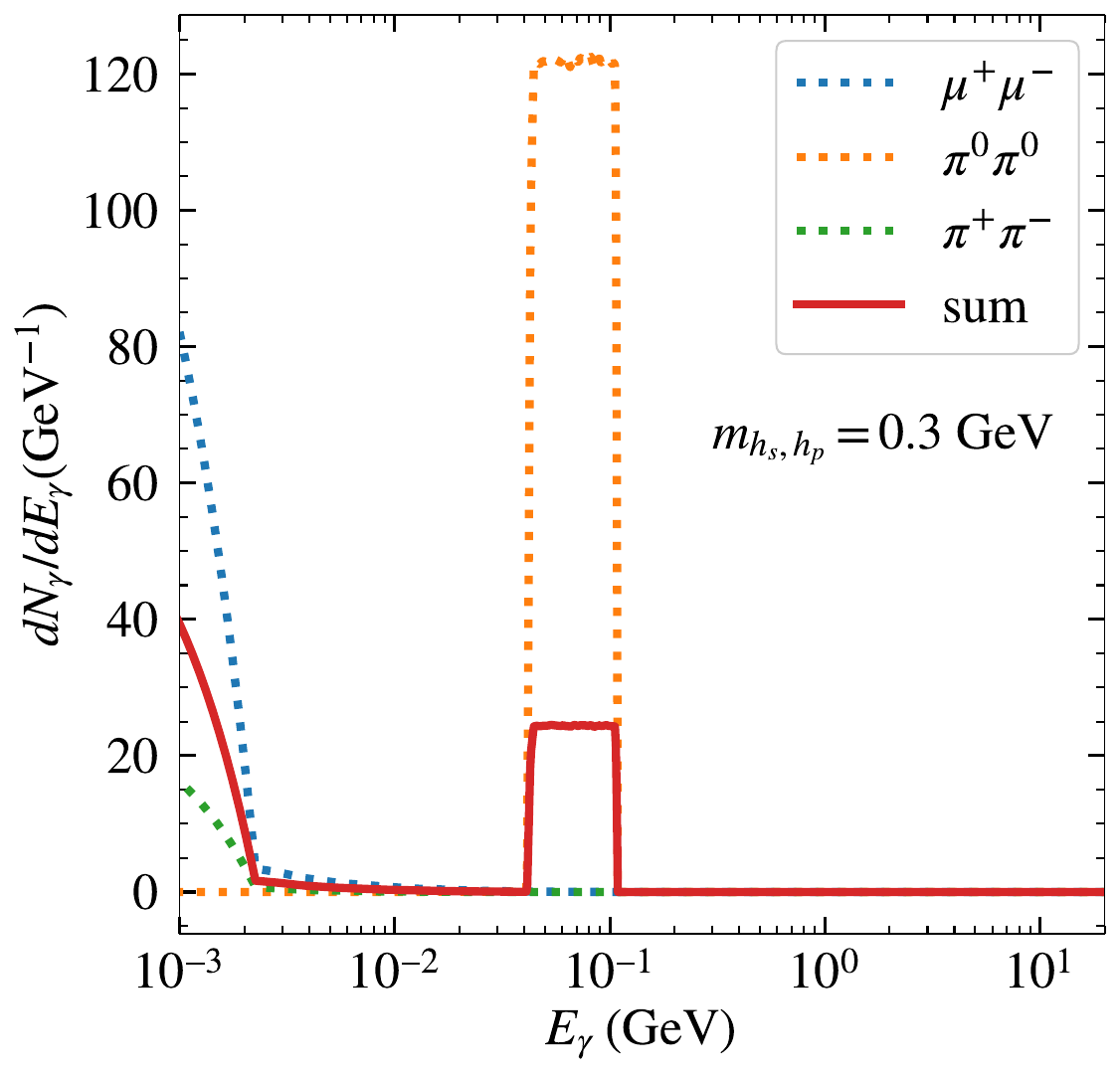}
 \caption{The photon spectrum in the mediator's center-of-mass frame generated from  \texttt{Pythia8} with $m_{h_s,h_p}=1\gev$ (Left panel) and  $m_{h_s,h_p}=0.3\gev$ (Right panel). The dotted lines represent the spectrum produced by each decay channel, and the solid lines represent the total spectrum added up according to their branching ratios.}
 \label{fig:rest_dnde}
\end{figure*}

After integrating the angle and centre velocity $v_{\rm CM}$, the one-dimensional velocity distribution can be expressed as a function of the relative velocity of DM particles $v_{\rm rel}$ and the position $r$~\cite{2018Anatomy}, 
\begin{eqnarray}
\label{eq:f_v_r_1D}
    f_{\rm ann.}[r, v_{\rm rel}]  =  \frac{v_{\rm rel}^2}{N_0} \times 
    &&\int_0^{v_{\rm esc}} dv_{\rm CM} v_{\rm CM}^2 
    \int_{-1}^1 d\cos\alpha
    \int_0^{2 \pi} d\phi \times \nonumber\\ 
    && \int_{-\mu_0}^{\mu_0} d\cos\theta \times
    f_1(v_{\rm rel}, v_{\rm CM} ,L_1,r)\times f_2(v_{\rm rel.}, v_{\rm CM},L_2,r), \\
    {\rm with}\quad 
    \mu_0  \equiv&&\frac{4 v_{\rm esc}^2- 4 v_{\rm CM}^2-v_{\rm rel}^2}{4 v_{\rm rel}v_{\rm CM}},
    \quad \cos\alpha \equiv  
    \frac{\mathbf{v_{\rm CM}}\cdot \mathbf{r}} {v_{\rm CM} \times r} \quad {\rm and} \quad v_{\rm esc}=\sqrt{2 \Phi_{\rm tot}(r)}.\nonumber
\end{eqnarray}
where $N_0$, $v_{\rm esc}$, $L_{1,2}$, $f_{1,2}$ and $\phi$ are the normalization factor, the escape velocity at $r_{\rm ann}$, angular momentum functions, relative velocity distribution function and azimuth angle, respectively~\footnote{The specific forms of $ L_{1,2} $ and $ f_{1,2} $, along with the detailed derivation process, can be found in Ref.~\cite{Lacroix:2018qqh}.}. 
Therefore, we take the velocity-averaged cross-section $\sv$ and photon-yield spectrum $\avgdnde$ for the DM annihilation at radius $r$ as
\begin{eqnarray}
    \sv[r] &=& \int \sigma v_{\rm rel}\times f_{\rm ann}[r, v_{\rm rel}] d v_{\rm rel}, ~~{\rm and}\nonumber\\
    \avgdnde &=& \int \dnde\times f_{\rm ann}[r, v_{\rm rel}] d v_{\rm rel}. 
\label{eq:avgXS}
\end{eqnarray}
 The photon spectrum $\dnde$ represents the photon energy distribution per DM annihilation. 
We employ \texttt{Pythia8}~\cite{Sjostrand:2007gs} to calculate the photon spectrum produced from light mediators in the center-of-mass frame and \texttt{CalcHEP}~\cite{Belanger:2013ywg} to calculate the annihilation cross-section, $\sigma v$. 
In this work, two benchmarks of mediator masses are set to calculate the spectrum, $m_{h_s,h_p} = 0.3$ GeV (the minimum mediator mass that can annihilate into $2\pi^0$) and $m_{h_s,h_p} = 1$ GeV (the maximum mediator mass considering $m_{\chi}\leqslant 10$ GeV with the secluded DM scenarios $m_{\chi} = 10 m_{h_s,h_p}$). We generate the spectra of light mediator to a pair of $\pi^0\pi^0$,  $\pi^{+}\pi^{-}$, and $\mu^{+}\mu^{-}$ in \texttt{Pythia8},
respectively, and then the spectra of individual processes are added up according to their decay branching ratios in Ref.~\cite{Chang:2016lfq}. Therefore, we can obtain the photon spectrum under the mediator rest frame with fixed $m_{h_s,h_p}$ as shown in Fig~\ref{fig:rest_dnde}. It's clear to see that only the process of $h_{s,p}\to \pi^0\pi^0$ is a box-like spectrum. 

It is worth mentioning that final-state radiation (FSR) is capable of producing somewhat distinctive features in the photon spectrum. FSR yields a continuous 1/E spectrum~\cite{Birkedal:2005ep}. The formula for the model-dependent FSR spectra of $\mu^{+}\mu^{-}$ and $\pi^{+}\pi^{-}$ final states are referenced from Eqs.~(C9) and~(C12) in Ref.~\cite{Liu:2014cma}: 
\begin{equation}
\begin{aligned}
  \frac{dN_\gamma}{dE_\gamma}\bigg|_{\pi^{+}\pi^{-}}
  = & \frac{\alpha}{2 \pi (r-1)^2(E-1)E} \bigg\{-[(-2+E)^2+4r(E-1)](r+E-1)  \nonumber \\
 & +(E-1)(-2r^2+2rE+E^2-2E+2) \rm ln\frac{1-E}{r}]\bigg\}
  \label{eq:FSR2-spectrum}
\end{aligned}
\end{equation}

\begin{equation}
\begin{aligned}
  \frac{dN_\gamma}{dE_\gamma}\bigg|_{\mu^{+}\mu^{-}}
  = & \frac{\alpha(1-E)}{3 \pi E} \bigg\{(3-2E+4E^{2}-2E^{3}) \rm ln \frac{1}{r}+[-\frac{17}{2}+\frac{23}{6}E-\frac{101}{12}E^{2}+\frac{55}{12}E^{3}) \nonumber \\
 & +(3-3E+4E^{2}-2E^{3}) \rm ln(1-E)]\bigg\},
  \label{eq:FSR1-spectrum}
\end{aligned}
\end{equation}
where $E$ is in the range of $ 0\leqslant E\leqslant (1-r)$ and $r=(\frac{m_{\mu}}{m_{\pi}})^2$. In our study, a pair of DM particles first annihilates into two mediators, which subsequently decay into SM final states. The resulting photon spectrum from these processes is distinctly different from that produced by DM directly annihilating into SM final states~\cite{DelaTorreLuque:2023olp,DallaValleGarcia:2024zva}. In addition, we present the spectra of the produced electrons, positrons, and neutrinos in the mediator rest frame in Appendix~\ref{sec:eve}.

\begin{figure*}[ht]
\centering
\includegraphics[width=7.8cm]{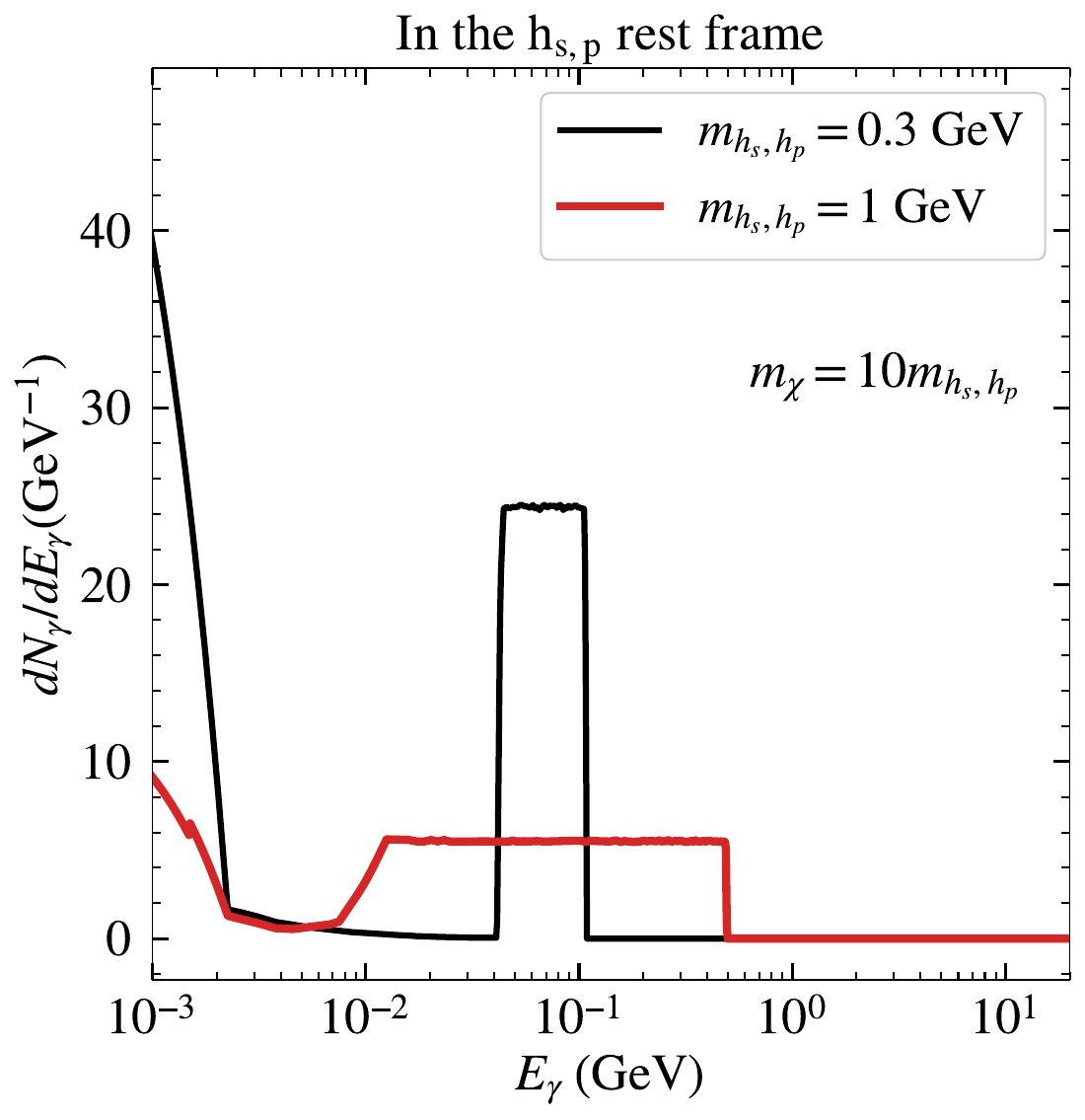}
\includegraphics[width=7.8cm]{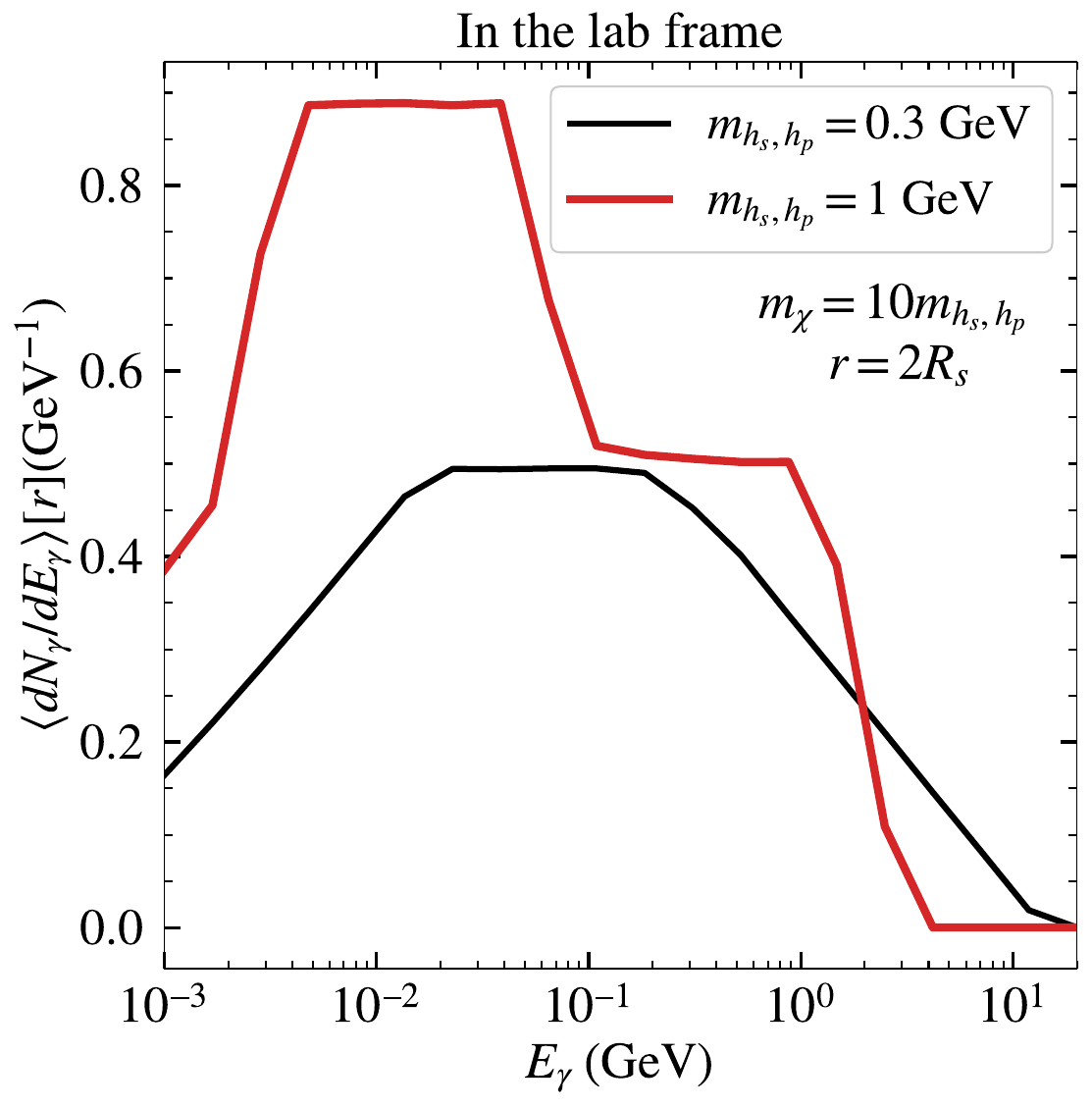}
 \caption{Left panel: the photon spectrum in the mediator's center-of-mass frame generated from \texttt{Pythia8}. Right panel: the velocity averaged photon spectra $\avgdnde$ in $p$-wave annihilation scenario which has been boosted in the lab frame for $m_{h_s,h_p}=0.3\gev$ (black line) and $m_{h_s,h_p}=1\gev$ (red line).}
 \label{fig:dnde}
\end{figure*}

Because low-energy photons produced from charged particles through FSR, the total photon spectral shape has a spike in the low-energy region as shown in the left panel of Fig~\ref{fig:dnde}. 
It is necessary to further make the Lorentz boost to the lab frame where the photon spectrum is observed. The final photon spectrum in the lab frame is a result of two-step boost. We first boost the photon spectrum in the mediator rest system to the DM pair center-of-mass frame and then to the lab frame, so this process is multiplied by two Lorentz factors: $\gamma_{h_{s,p}}$, $\gamma_c$. Here $\gamma_{h_{s,p}} \equiv 1 / \sqrt{1 - v^2_{h_{s,p}}}$,
$v_{h_{s,p}} \equiv \sqrt{1 - 4 m^2_{h_{s,p}} / s}$ and $\gamma_{c} \equiv 1 / \sqrt{1 - v^2_{\rm CM}}$.

On the other hand, the DM annihilation process that generates a box-shaped photon spectrum can be represented by the following equation~\cite{Boddy:2015efa}: 
\begin{equation}
  \frac{dN_\gamma}{dE_\gamma}
  =\sum_{i} \frac{S_i}{\Delta E_i} \left[\Theta(E-E_-)-\Theta(E - E_+)\right]\ ,
  \label{eq:box-spectrum}
\end{equation}
where the $i$ is the $i$-th bin, $E_\pm$ are the kinematic edges, $E_\pm \equiv E_\pm^{i}\gamma_{h_{s,p}}(1\pm v_{h_{s,p}})\times \gamma_{c}(1\pm v_c) $ and $\Delta E \equiv E_+^{i} - E_-^{i}$ is the box width.
Subsequently, the spectrum obtained from Pythia8 is divided into $100$ bins\footnote{We have tested various binning schemes and found that increasing the number of bins beyond $80$ resulted in negligible differences in the results; hence, we opted for $100$ bins in this work.}. The small and large energy boundaries of each bin are labeled as $E_-^{i}$,$E_+^{i}$, and the area of each small bin is labeled as $S_i$. These $100$ bins are substituted into Eq.~(\ref{eq:box-spectrum}) for numerical calculation and then added together. Finally, the photon spectrum in the lab frame can be obtained as shown in the right panel of Fig~\ref{fig:dnde}.

\subsection{P-wave annihilation scenario}
\label{sec:pwave}

\begin{figure*}[ht]
\centering
\includegraphics[width=7.8cm]{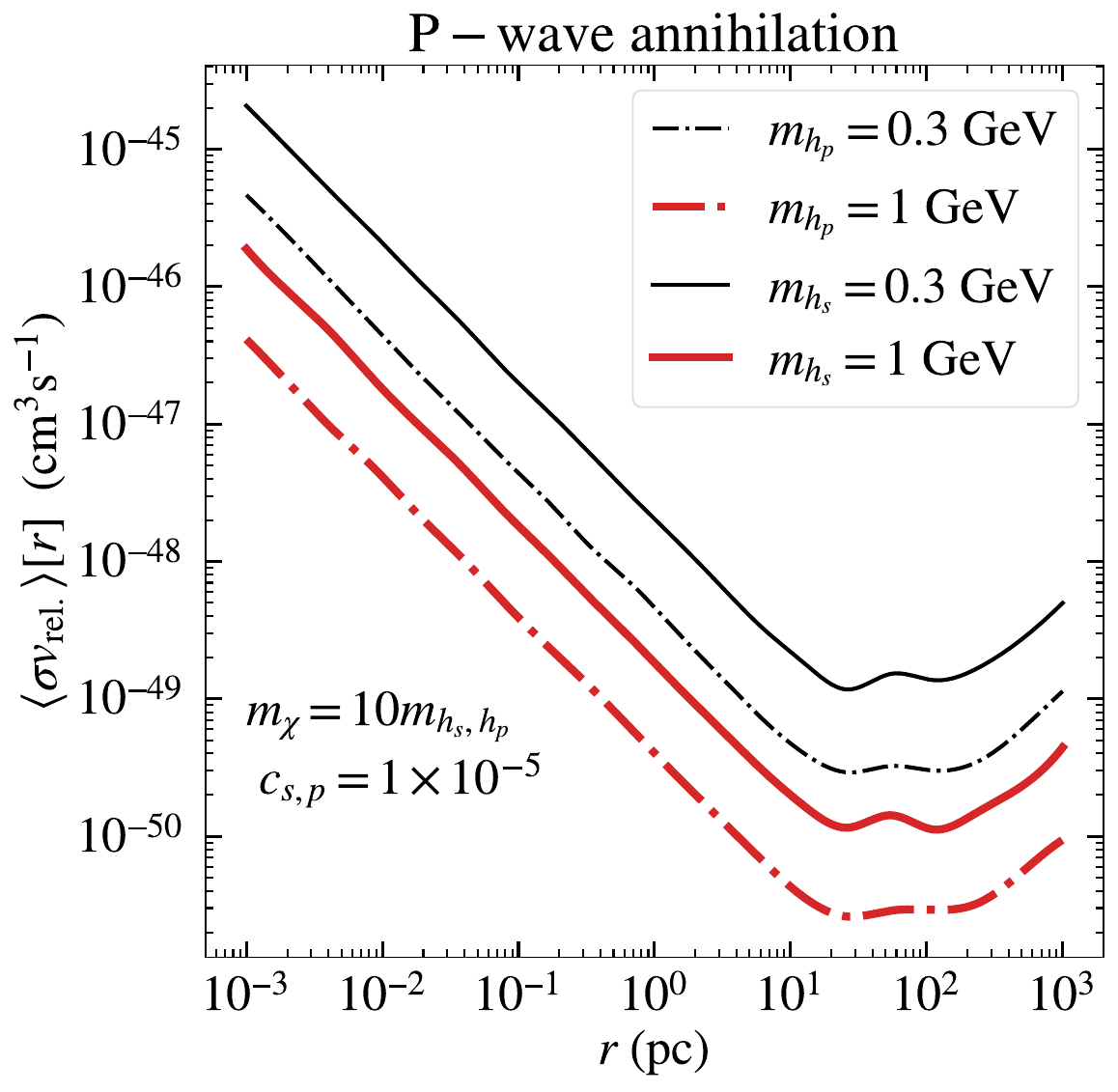}
 \caption{The radial distribution of velocity-averaged cross-section $\sv$. The dashed lines represent pseduoscalar mediator case and the solid lines represent scalar mediator case for $m_{h_s,h_p}=0.3\gev$ (black line) and $m_{h_s,h_p}=1\gev$ (red line), respectively.}
 \label{fig:pw_sigmv}
\end{figure*}

We first consider the cross-sections for DM annihilation into a pair of scalar or pseudoscalar mediators. Based on the interactions in Eq.~\eqref{eq:DM-Phi}, the cross-sections $\sigma(\chi\bar{\chi}\to h_s h_s)$ and $\sigma(\chi\bar{\chi}\to h_p h_p)$ can be expressed as  
\begin{eqnarray}\label{eq:cs}
\sigma(\chi\bar{\chi}\rightarrow h_s h_s) &=& \frac{(c_{s}/2)^4}{32 \pi R_\chi^2 s^2}\left( 
\frac{6 m_{h_s}^4 - 4 m_{h_s}^2 (4 m_{\chi}^2 + s) - 32 m_{\chi}^4 + 16 m_{\chi}^2 s + s^2}{2 m_{h_s}^2 - s} \right. \nonumber \\
&& \left. \times \ln\left(\frac{-R_{h_s} R_\chi s - 2 m_{h_s}^2 + s}{R_{h_s} R_\chi s - 2 m_{h_s}^2 + s}\right) \right. \nonumber \\
&& \left. - \frac{ R_{h_s} R_\chi s \left( 3 m_{h_s}^4 - 16 m_{h_s}^2 m_{\chi}^2 + 2 m_{\chi}^2 (8 m_{\chi}^2 + s) \right)}{m_{h_s}^4 - 4 m_{h_s}^2 m_{\chi}^2 + m_{\chi}^2 s} \right), 
\end{eqnarray}

\begin{align}\label{eq:cp}
\nonumber\sigma(\chi\bar{\chi}\rightarrow {h_p} {h_p}) &= \frac{(c_{p}/2)^4}{16\pi s} \Bigg[ \frac{6m_{h_p}^4 - 4m_{h_p}^2 s + s^2}{(s - 2m_{h_p}^2)(s - 4m_\chi^2)} \ln \tfrac{R_{h_p}^2 + 1 + 2 R_{h_p} R_\chi}{R_{h_p}^2 + 1 - 2 R_{h_p} R_\chi} \nonumber \\
& - 2\frac{R_{h_p}}{R_\chi} - \frac{m_{h_p}^4}{m_{h_p}^4 - 4 m_{h_p}^2 m_\chi^2 + m_\chi^2 s} \frac{R_{h_p}}{R_\chi} \Bigg],
\end{align}
where $R_\chi=\sqrt{1-\frac{4m_\chi^2}{s}}$ and $R_{h_p, h_s }=\sqrt{1-\frac{4m_{h_p, h_s}^2}{s}}$ as the velocity of DM in the initial state and mediator in the center-of mass frame. 
From Eq.~(B5) in the appendix of Ref.~\cite{Cannoni:2015wba}, we can derive the relationship between center-of-mass energy and relative velocity $s = 4 m_\chi^2 \left( \frac{4}{4 - v_{\text{rel}}^2} \right)$. By substitute the relative velocity $v_{\rm rel}$ into the above expression for the cross-section, we find that the cross-section satisfies $\sigma v \propto v_{\text{rel}}^2$, indicating a p-wave annihilation. 
Using the relative velocity $v_{\rm rel}=2R_{\chi}$, we then perform the integral of velocities to obtain $\sv$ by Eq.~\eqref{eq:avgXS}. In Fig.~\ref{fig:pw_sigmv}, it shows results of $\sv$ with $m_{\chi}=10m_{h_s,h_p}$ and $c_{s,p}=10^{-5}$ for $p$-wave annihilation with two benchmark points $m_{h_p, h_s} = 0.3$ and $1$ GeV. 
Because the cross-section formula with the scalar mediator includes an additional factor of $1/R_{\chi}^2$, where $ R_{\chi} < 1 $, the DM annihilation cross-section with the scalar mediator is slightly greater than that with the pseudoscalar mediator. 
For $r<10^2 \pc$, because $\Phi\propto r^{-1}$ or 
$v_{\rm esc} \propto r^{-1/2}$ where $v_{\rm esc}$ increases with the radius, hence $\sigma v_{\rm rel.} \propto r^{-1}$, while it increases with radius for $r>10^2 \pc$. 

\subsection{Forbidden annihilation scenario}
\label{sec:forbidden}
In most cases, the forbidden annihilation channel is inactive in the contemporary Universe, as DM typically annihilates into lighter particles because of the kinematic threshold. However, the acceleration from SMBH has the potential to reactivate the forbidden annihilation channel $ \chi\bar{\chi}\rightarrow h_{s,p}h_{s,p} $ with $m_{\chi}\lesssim m_{h_s,h_p}$. Given that their $\langle \sigma v_{\text{rel.}} \rangle$ are constrained by limited phase space, an enhanced $ c_{s,p} $ compared to $p$-wave annihilation scenarios might be necessary to meet the detection thresholds of gamma-ray telescopes.

\begin{figure*}[ht]
\centering
\includegraphics[width=7.7cm]{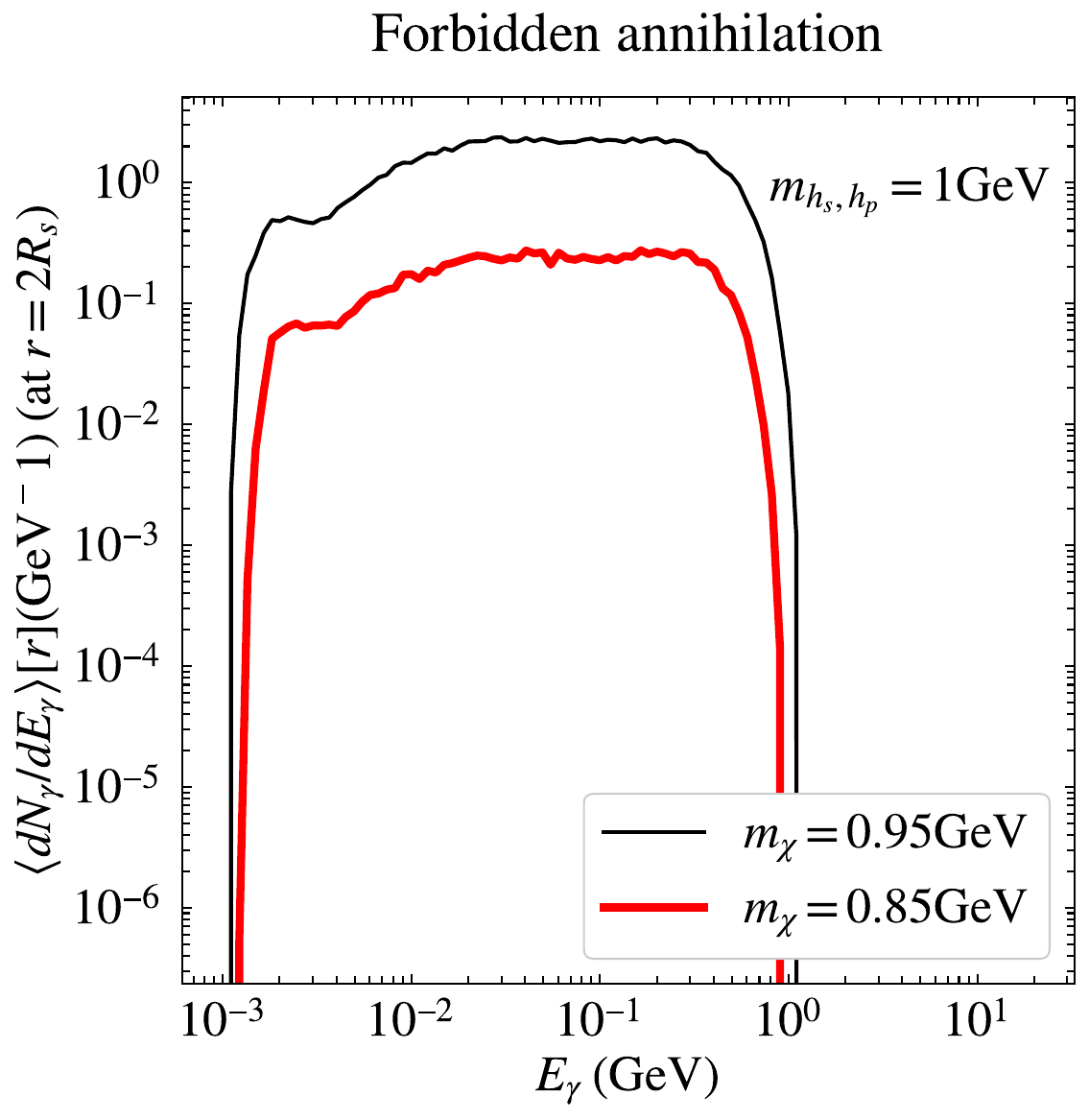}
\includegraphics[width=7.9cm]{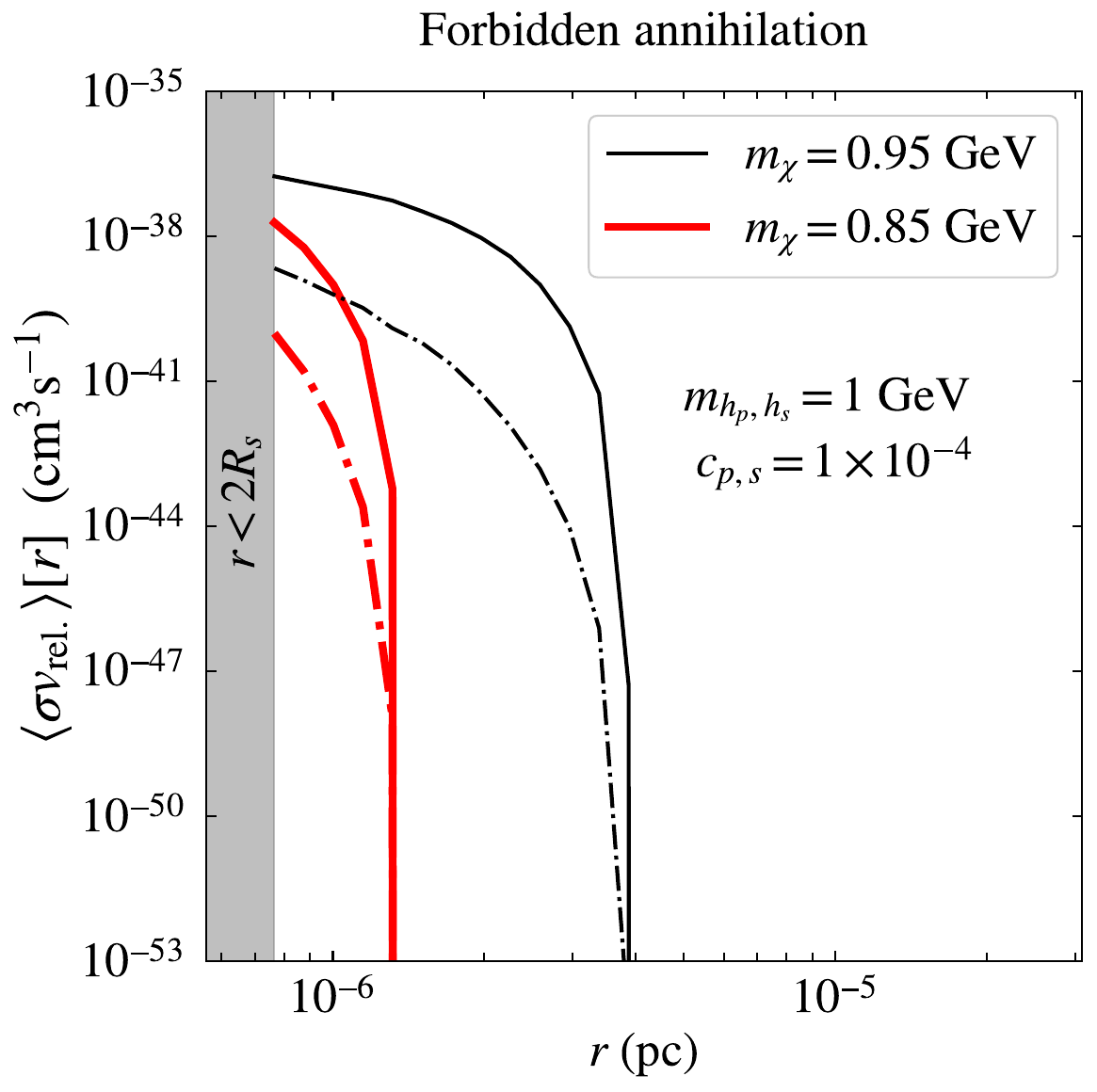}
 \caption{Left panel: the velocity averaged photon energy distribution at $r=2R_s$ for $m_{h_s,h_p}=1\gev$ with $m_{\chi}=0.95\gev$ (black line) and $m_{\chi}=0.85\gev$ (red line), respectively. Right panel: the radial distribution of $\sv$ in the forbidden annihilation scenario as a function of $r$ for $c_{s,p}=10^{-4}$. Here solid lines are scalar mediator scenario and dashed lines are pseduoscalar mediator scenarios.}
 \label{fig:forb}
\end{figure*}

In the left panel of Fig.~\ref{fig:forb},  we showcase the broadest $\avgdnde$ for the forbidden annihilation scenario at a radius of $r=2 R_s$. This is because the acceleration effect of the black hole intensifies closer to its center, causing forbidden annihilation channel at the capture radius of $2R_s$ to yield the highest photon count and the broadest spectrum.  We also examine $\avgdnde$ values for $m_{\chi}=0.95\gev$ (black line) and  $m_{\chi}=0.85\gev$ (red line) with $m_{h_s,h_p}=1\gev$. Our results indicate that different relationships between the DM mass $m_{\chi}$ and mediator mass $m_{h_s,h_p}$ yield distinct spectra. The smaller the mass difference between them, the lower the energy required to open the forbidden annihilation channels, resulting in a greater number of produced photons and thus a larger flux (as shown in the right panel of Fig.~\ref{fig:flux}). Notably, the breadth of the spectrum is unaffected by the choices of coupling $c_{s,p}$.

In the right panel of Fig.~\ref{fig:forb}, employing the same benchmark points, we compare their cross-sections $\sv$ with $c_{s,p}=10^{-4}$.  We observe that a larger mass difference between $m_{h_s,h_p}$ and $m_{\chi}$, complicates the activation of the forbidden annihilation channel due to the kinematic threshold, necessitating more significant acceleration from the black hole. Consequently, with a greater mass difference (red lines), the cross-section profile approaches closer to the capture radius. Moreover, akin to the $p$-wave annihilation scenario, in the forbidden annihilation scenario, the cross-section of the scalar mediator (solid lines) surpasses that of the pseudoscalar mediator (dashed lines).

\section{GAMMA-RAY FLUX FROM GC}
\label{sec:GC}

\begin{figure*}[ht]
\centering
\includegraphics[width=7.8cm]{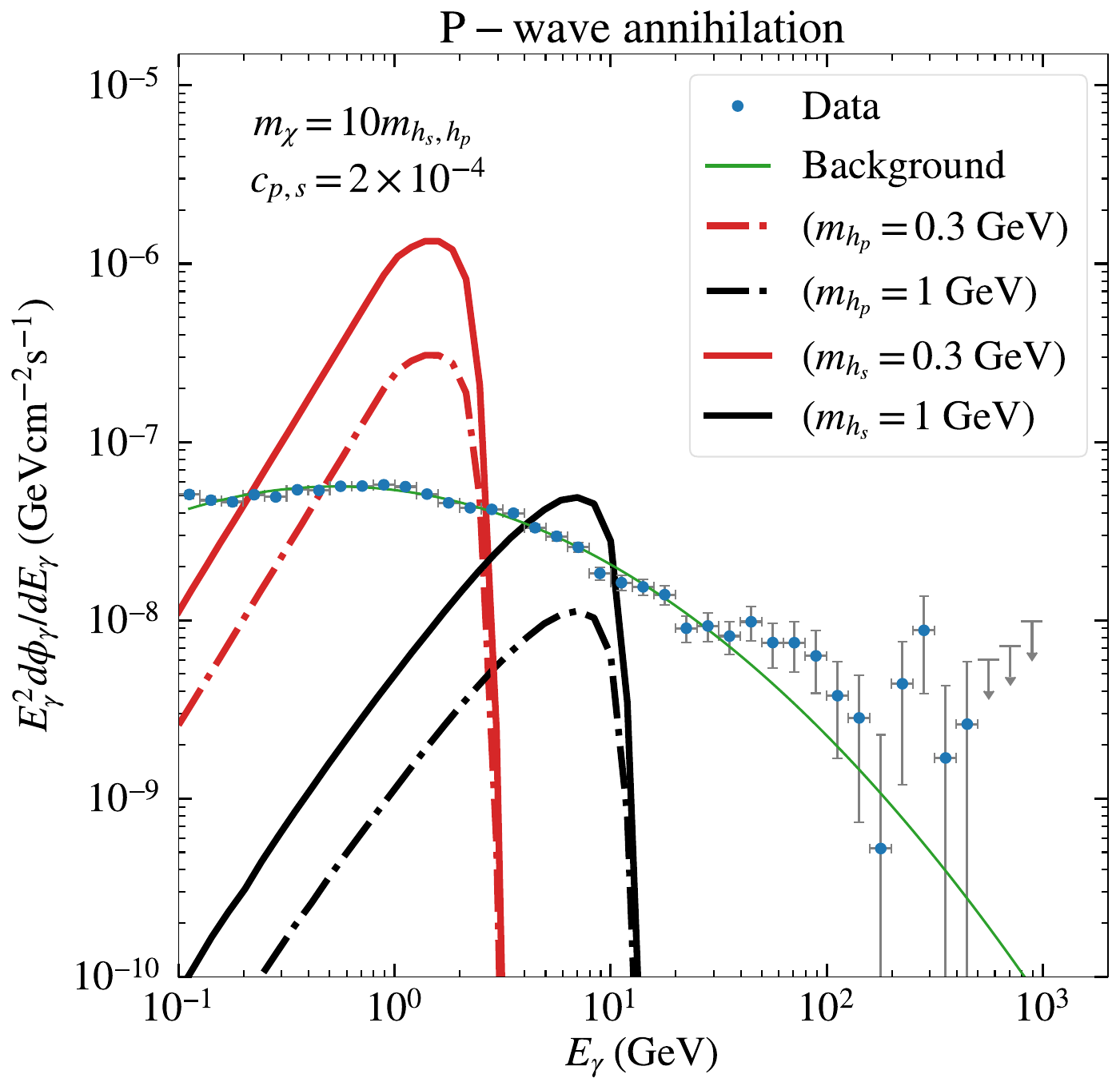}
\includegraphics[width=7.8cm]{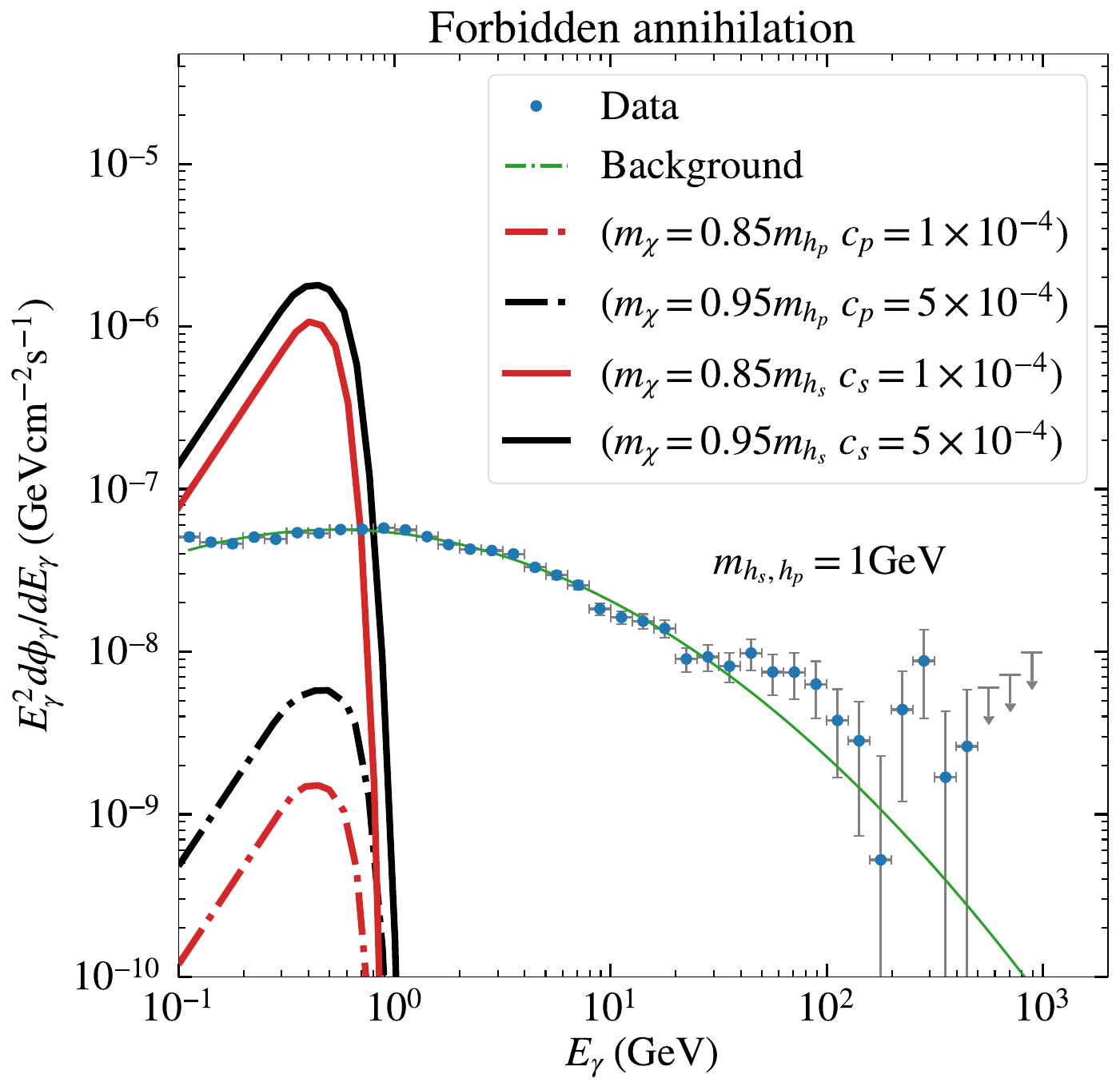}
 \caption{The expected photon energy spectra of DM signal are presented for $p$-wave annihilation (left panel) and forbidden annihilation (right panel) with various benchmark points. Blue circles depict the Fermi-LAT measured spectral energy distributions (SEDs) of Sgr A$^{\star}$, and the green line represents the background. }
 \label{fig:flux}
\end{figure*}

The vicinity of the GC is an ideal region for detecting photon signals produced by DM annihilation. After obtaining the spectrum $\avgdnde$ and the DM density profile $\rho_\chi(r) $, the photon flux from DM annihilation near Sgr A$^{\star}$, as observed on Earth, can be expressed as follows:
\begin{equation}
\label{eq:flux}
\begin{aligned}
    \frac{d\phi_{\gamma}}{dE_{\gamma}}= \frac{1}{4\pi m_{\chi}^{2}D_{\odot}^{2}} &\times \int_{2R_{\rm s}}^{0.4\kpc} \rho^2_\chi(r) r^2 dr
    \int_0^{2v_{\rm esc}} \frac{dN_{\gamma}}{dE_{\gamma}}\sigma v_{\rm rel.} f_{\rm ann}d{v}_{\rm rel.},
\end{aligned}
\end{equation}
with the distance $D_{\odot}=8.5\kpc$ from Earth to Sgr A$^{\star}$. In this study, Sgr A$^{\star}$ is considered as a point source, enabling us to integrate $r$ up to a distance of $0.4\kpc$. The maximum DM velocity near the SMBH at the GC is given by the escape velocity $ v_{\text{esc}} $. Considering a collision between two DM particles with equal speeds in opposite directions, the maximum relative velocity $v_{\rm rel}$ is $ 2 v_{\text{esc}} $. Hence, we set the upper limit for the integration of the relative velocity to $ 2 v_{\text{esc}} $.
Fig.~\ref{fig:flux} displays the predicted gamma-ray spectra for both the $p$-wave and forbidden scenarios. The source of the spectral data used here, as well as the processing methods, will be described in details in the Sec.~\ref{sec:data}. 

In the $ p $-wave scenario (left panel of Fig.~\ref{fig:flux}), we set the typical secluded DM relation, $m_{\chi} = 10 m_{h_s,h_p}$. For ease of presentation, we choose the parameters $c_{p,s}=1.5\times10^{-4}$ and $m_{hs,hp}=0.3 \text{GeV}$ as the benchmark points. The photon spectrum represented by the black lines ($m_{hs,hp}=1\gev$) is less intense than that depicted by red lines ($m_{hs,hp}=0.3\gev$). In the scalar mediator case (solid lines), the flux is approximately double that observed in the pseudoscalar mediator case (dashed lines) with the same setting of parameters. The smaller the mediator mass, the greater the contribution from FSR, resulting in a higher number of low-energy photons and a peak in the spectral profile that shifts towards the low-energy region. Therefore, the flux represented by the red lines ($ m_{h_s, h_p} = 0.3 \, \text{GeV} $) is greater than that of the black lines ($ m_{h_s, h_p} = 1 \, \text{GeV} $) and its peak is closer to the low-energy region. Additionally, in the scalar mediator case (solid lines), the DM annihilation cross-section is larger than that in the pseudoscalar mediator case, leading to a similar relationship in their fluxes. Specifically, the flux at the peak in the scalar mediator case is approximately twice that of the flux observed in the pseudoscalar mediator case (dashed lines). 

In the forbidden scenario (right panel of Fig.~\ref{fig:flux}), since the vertical axis represents the flux multiplied by $ E^2 $, the curve becomes increasingly steep with the increase in energy $ E $. Subsequently, due to the conservation of energy, the lines begin to decrease gradually as it approaches the mass of the mediator, ultimately truncating around the photon energy of $1$ GeV (mass of the mediator).

\section{ANALYSIS AND RESULT}
\label{sec:data}
\subsection{Data analysis}

We use nearly 16 years of Fermi-LAT P8R3 data within the energy range ($100$ MeV $-$ $1$ TeV) from October 27, 2008\footnote{The photon (above $30$ GeV) data before October 27 have a significantly higher level of background contamination as seen in \url{https://fermi.gsfc.nasa.gov/ssc/data/analysis/LAT_caveats.html}.}, to August 18, 2024~\cite{Fermi-LAT:2021wbg}.
Here the photon events in 10$^\circ$ of Sgr A$^{\star}$ with the {\tt SOURCE} event class and {\tt FRONT+BACK} conversion type are selected in our analysis and events with zenith angles larger than 90$^\circ$ are excluded to reduce the contamination from the Earth’s limb.
Then we extract good time intervals with the quality-filter cut {\tt (DATA\_QUAL==1 \&\& LAT\_CONFIG==1)}.
Here we divide the selected data into 40 evenly spaced logarithmic energy bins from $100$ MeV to $1$ TeV.
To calculate the spectral energy distribution (SED) of Sgr A$^{\star}$, the standard binned likelihood analysis\footnote{\url{https://fermi.gsfc.nasa.gov/ssc/data/analysis/scitools/binned_likelihood_tutorial.html}} are performed in each energy bin.\footnote{Other analysis details can be found in our previous work~\cite{Yang:2024jtp}. }
The SEDs of Sgr A$^{\star}$ we obtained are plotted in the Fig.~\ref{fig:flux}.

In the Fourth Fermi-LAT source catalog~\citep{2020AbdollahiApJS} (4FGL), Sgr A$^{\star}$ have been identified as the point source 4FGL J1745.6-2859 and shown significant gamma-ray emissions of astrophysical origin.
To seek out potential DM signals above this astrophysical background, we refit the measured SEDs with the null hypothesis model ($H{_0}$, DM signal absent) and the alternative hypothesis model ($H{_1}$, DM signal present), respectively.
For the $H{_0}$ model, we simply model the gamma-ray emission from the astrophysical background as the log-parabola function corresponding to the recommended spectral model for 4FGL J1745.6-2859 in the 4FGL:
\begin{align}
\label{logp}
  \frac{d \phi}{d E} &= N_0 \left( \frac{E}{E_{b}} \right)^{-\alpha - \beta \ln (E / E_{b})},  
\end{align}
where the scale parameter $E_{b}$ is fixed at 6499.68 MeV as default in 4FGL; $N_0$, $\alpha$ and $\beta$ are set as free parameters in the refitting process.
As for the $H{_1}$ model, we add the expected flux of DM signal given in Eq.~\ref{eq:flux} to the $H{_0}$ model.
Refitting SEDs with the $H_0$ or $H_1$ model, we utilize the $\chi^2$ statistical method: $\chi^2$=$\sum$ $\frac{[M_i({\bf \theta})-D_i]^2}{\sigma_i^2}$,
where $M_i({\bf \theta})$ is expected flux for the hypothesis model in $i$th energy bin, ${\bf \theta}$ represents a set of free parameters, $D_i$ and $\sigma_i$ are measured flux and corresponding error in $i$th energy bin.
In order to evaluate the significance of DM signals, we establish a test statistic (TS) defined by $ {\rm TS} = {\chi^2_{H_0} - \chi^2_{H_1}}$, which follows the $\chi^2$ distribution with 2 degrees of freedom.
To further constrain the DM properties, we also derive the 95\% confidence level upper limits at where the $\chi^2_{H_1}$ value exceeds the $\chi^2_{H_0}$ value by $6.18$.

\subsection{Result}

\begin{figure*}[ht]
\centering
\includegraphics[width=8cm]{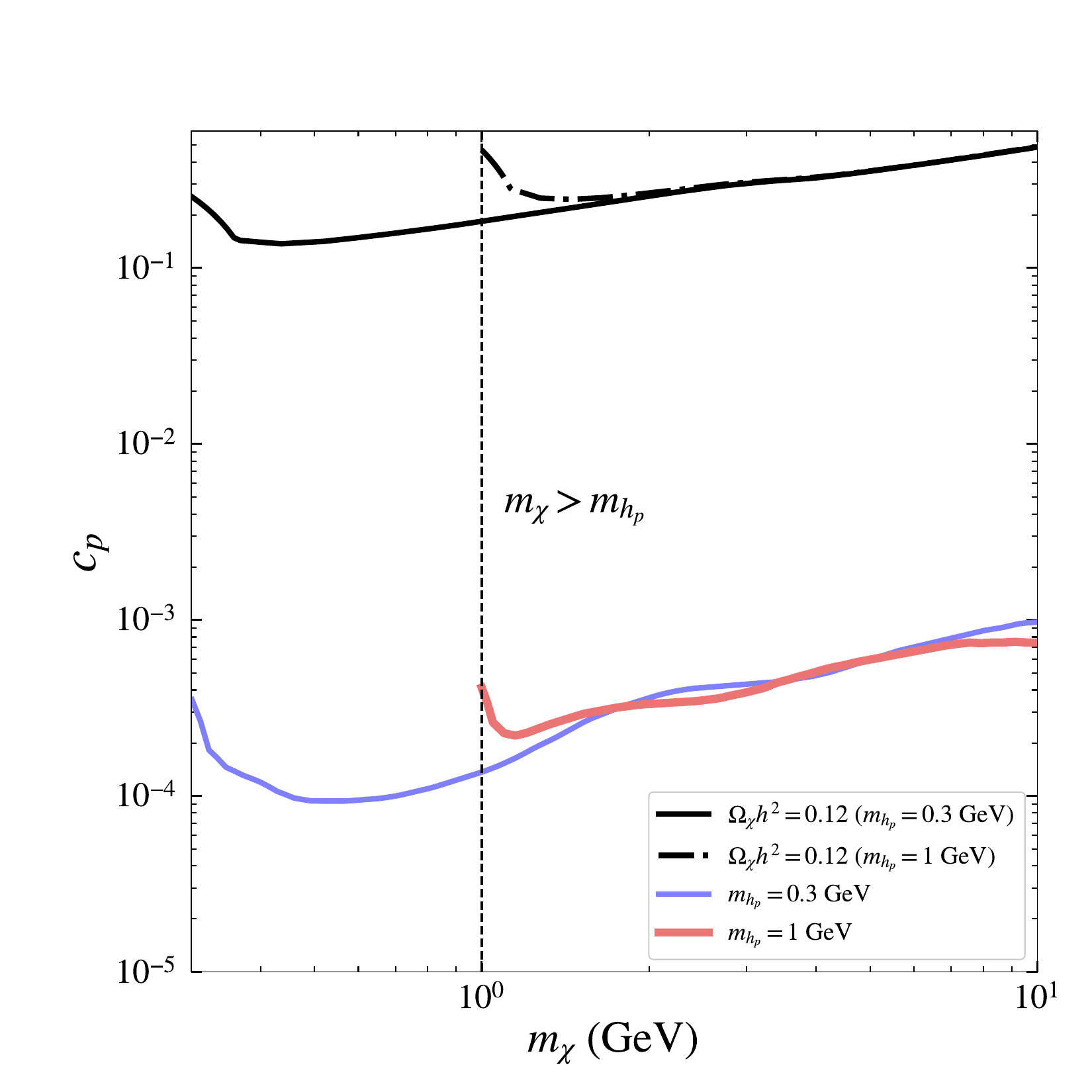}
\includegraphics[width=8cm]{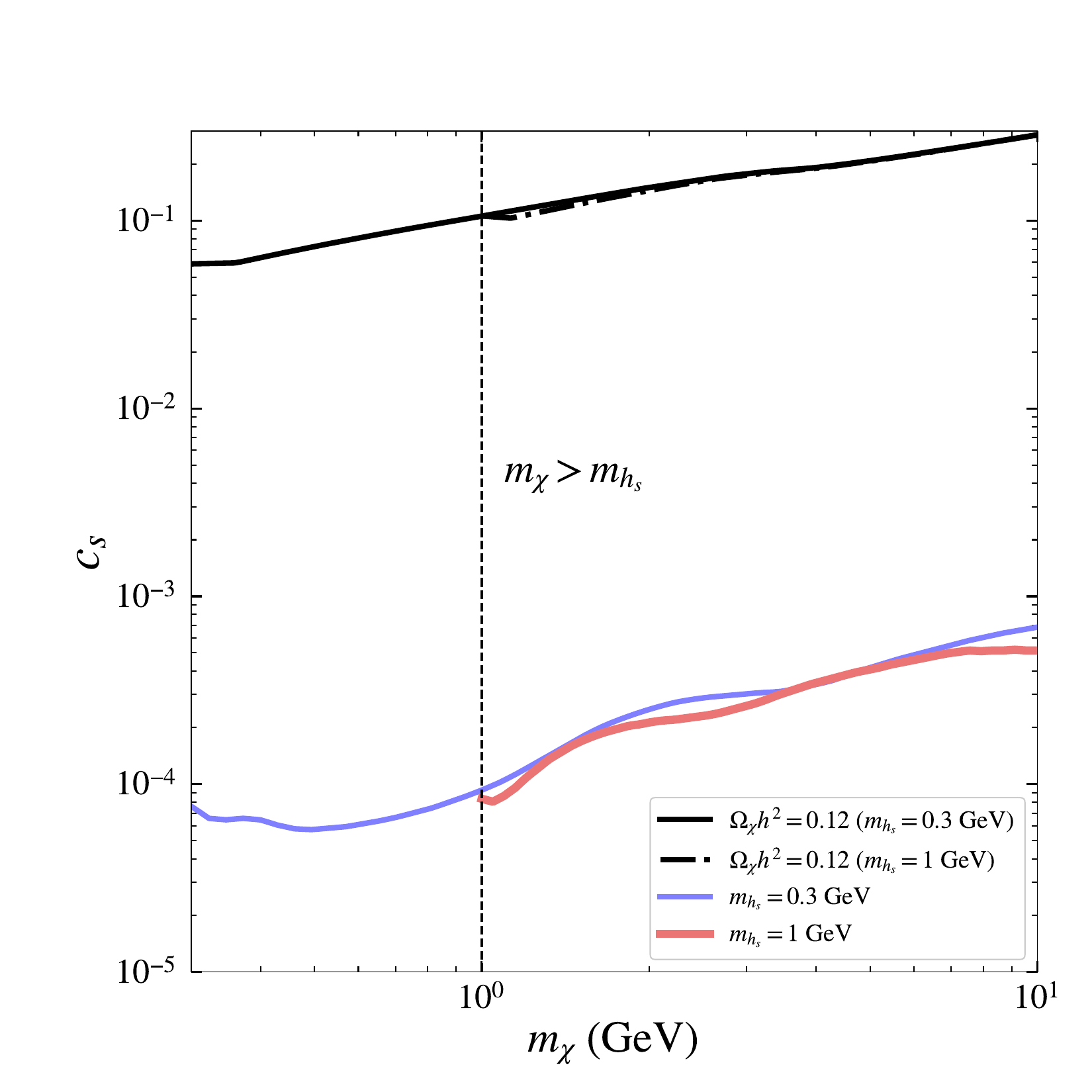}
\includegraphics[width=8cm]{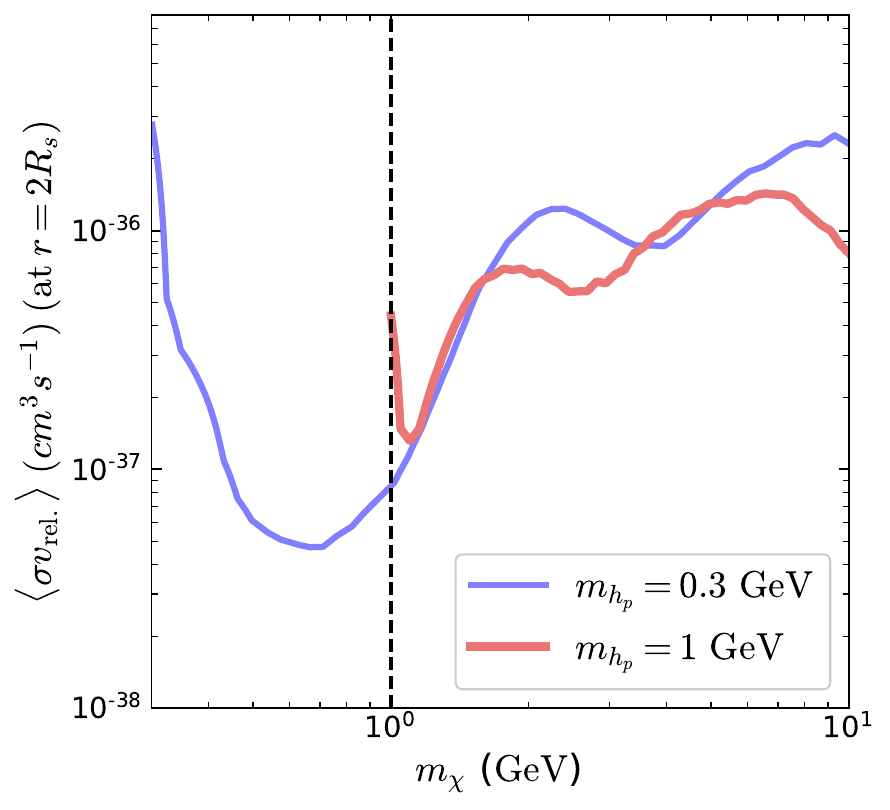}
\includegraphics[width=8cm]{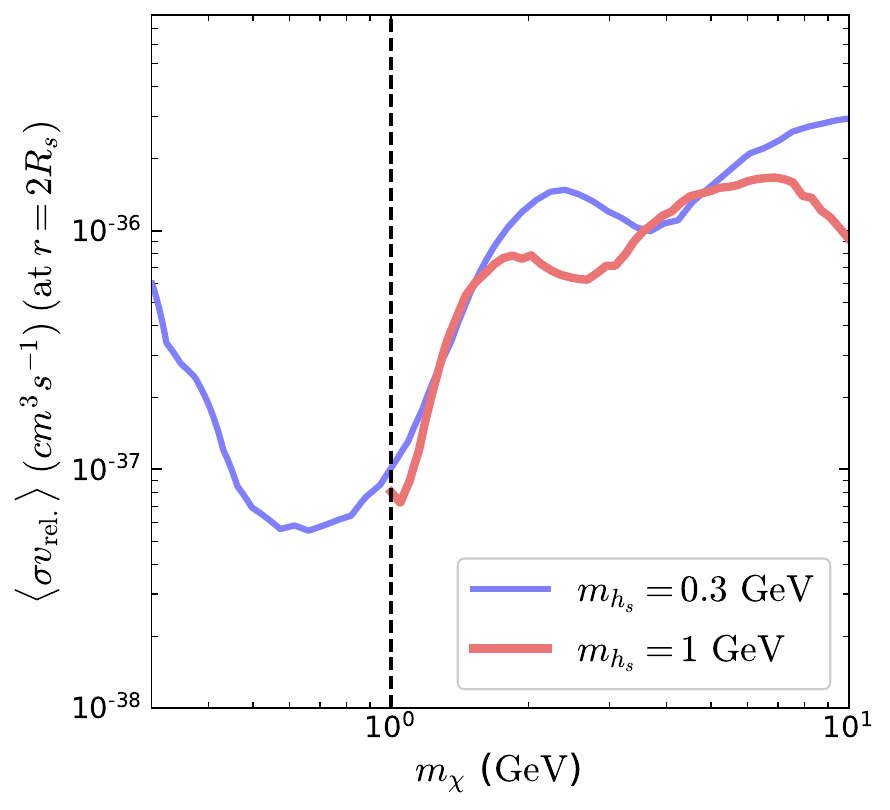}
 \caption{The $95\%$ upper limits for $c_{p}$ (left panels), $c_{s}$ (right panels) and $\sv$ at $r=2R_s$ (bottom panels), based on Fermi-LAT continuum spectrum analysis in scenarios where $m_{\chi} > m_{h_s,h_p}$.}
 \label{fig:fermi_con}
\end{figure*}

Due to limited data availability in the forbidden annihilation scenario, we focus solely on the $p$-wave annihilation scenario to constrain the coupling constant $c_{s,p}$ and the average relative velocity cross-section \sv. 
Using Fermi-LAT measured SED of  Sgr A$^{\star}$, we search for possible DM signals and futher set the $95\%$ confidence level upper limits on DM parameters  for two distinct annihilation processes $\chi\bar{\chi}\to h_p h_p$ and $\chi\bar{\chi}\to h_s h_s$, for the DM mass range of $0.3$ GeV $<m_{\chi}<$ $10$ GeV, which are depicted as follows:
\begin{itemize}
\item For $\chi\bar{\chi}\to h_p h_p$ process, in the case of $ m_{h_p} = 1 \gev$ (0.3 GeV),  we find best-fit DM parameters  $m_\chi/(\gev)$,$c_{p}$, $\sv/({\rm cm}^3/{\rm s}) =1.68, 2.53\times 10^{-4}, 2.94\times 10^{-37} \, (0.30, 3.02\times 10^{-4}, 1.41\times 10^{-36})$ have a TS value of 7.78 (13.55)
and a corresponding significance level of 2.32$\sigma$ (3.25$\sigma$).
Considering relatively poor spatial resolution of Fermi-LAT at low energies and potential contamination from the intricate radiation background surrounding Sgr A$^\star$, coupled with the suspected signal failing to exceed the 5$\sigma$ significance threshold, we conservatively choose to further constrain the DM parameter space.
As shown in the left two panels of Fig.~\ref{fig:fermi_con}, when $ m_{h_p} = 1 \gev$ (red line), the upper limits on the cross-section $\sv/({\rm cm}^3/{\rm s})$ at $r = 2R_s$ range from $1.4 \times 10^{-37} $ to $ 1.5 \times 10^{-36} $, while the upper limits on the coupling constant $c_p$ span from $ 2.0 \times 10^{-4} $ to $ 7.7 \times 10^{-4} $. For thermal freeze-out mechanism, the relic abundance, $\Omega_{\chi}h^2=0.12$~\cite{Planck:2018vyg}, obtained by micrOMGAs~\cite{Belanger:2013ywg} predicts the coupling constant of $ 0.129 $ to $ 0.487 $ (dashed black line). In contrast, when $ m_{h_p} = 0.3 \gev$ (blue line), the upper limits on the cross-section $\sv/({\rm cm}^3/{\rm s})$ extend from $ 4.6 \times 10^{-38} $ to $ 2.6 \times 10^{-36} $, and the upper limits on the coupling constant $c_p$ range from $ 8.6 \times 10^{-5} $ to $ 9.5 \times 10^{-4} $. The relic abundance additionally predicts the coupling constant from $ 0.248 $ to $ 0.487 $ (solid black line). 
The constraints on the coupling constant $c_{p}$ are significantly stronger than those required by the freeze-out mechanism, whereas DM produced via the freeze-in mechanism necessitates coupling constants that are substantially smaller than those in the freeze-out scenario. In Ref.~\cite{Bharucha:2022lty}, the authors studied a similar DM model to ours and indicated that if DM is produced through the freeze-in mechanism, the coupling constant values range from $ 10^{-7} $ to $ 10^{-13} $ for reheating temperatures of $ T_{\text{RH}} = 200 - 2000 \, \text{GeV} $ and mass ratios $m_{\chi}/m_{h_p}=10$, which correspond to the mass range we consider, as shown in Fig.~5 of their paper. Consequently, there is a significant gap between the predicted constraints on the coupling constants from our analysis and the expected values from the freeze-in mechanism.

\item 
 For $\chi\bar{\chi}\to {h_s} {h_s}$ process, when considering the scenario of $m_{h_s}$ equals 1 GeV (0.3 GeV) we have obtained the best-fit DM parameters  $m_\chi/(\gev)$, $c_s$, $\sv/({\rm cm}^3/{\rm s}) = 1.68, 1.55\times 10^{-4}, 3.21\times 10^{-37} \, (0.30, 6.3\times 10^{-5} , 2.83\times 10^{-37})$, which yield a TS value of 7.77 (14.28), corresponding to a significance level of 2.32$\sigma$ (3.35$\sigma$).
 We display the $95\%$ confidence level upper limits on DM parameters for the $\chi\bar{\chi}\to {h_s} {h_s}$ process in the right two panels of Fig.~\ref{fig:fermi_con}.
 In the case of $ m_{h_s} = 1 \gev$ (red line), the upper limits on the cross-section $\sv/({\rm cm}^3/{\rm s})$ range from $ 6.8 \times 10^{-38} $ to $ 1.7 \times 10^{-36} $, and the upper limits on the coupling constant $c_s$ vary from $ 7.8 \times 10^{-5} $ to $ 5.1 \times 10^{-4} $. For thermal freeze-out mechanism, the relic abundance predicts the coupling constant of the range $ 0.101 $ to $ 0.289 $. Conversely, when $ m_{h_s} = 0.3 $$\gev$ (blue line), the upper limits on the cross-section $\sv/({\rm cm}^3/{\rm s})$ are from $ 5.4 \times 10^{-38} $ to $ 2.9 \times 10^{-36} $, and the upper limits on the coupling constant $c_s$ range from $ 5.5 \times 10^{-5} $ to $ 6.9 \times 10^{-4} $. The relic abundance predicts the coupling constant extending from $ 0.058 $ to $ 0.289 $ (solid black line).
\end{itemize}

\section{SUMMARY AND DISCUSSION}
\label{sec:sum}

For ${\cal O}(1)$ GeV dark matter (DM) candidates, direct detection and accelerator experiments have already imposed stringent constraints on the coupling between DM particles and Standard Model (SM) particles. In this framework, the secluded DM scenario presents itself as a compelling alternative. In this scenario, the DM annihilation cross-section is not directly linked to the coupling between mediators and SM particles. Consequently, the coupling between mediators and SM particles can be much smaller, rendering direct detection and accelerator experiments less effective in searching for this type of DM. Instead, indirect detection becomes a powerful approach for studying secluded DM, particularly for probing the interactions between DM particles and mediators. For light DM with masses below $10$ GeV, the gravitational acceleration effects near the supermassive black hole at the Galactic Center (GC) allow for effective detection of DM $ p $-wave and forbidden annihilation processes, thereby increasing detection efficiency. 
We display the expected photon energy spectra of DM signals for $ p $-wave annihilation and forbidden annihilation in the Fig.~\ref{fig:flux}.
Due to the limited parameter space for forbidden annihilation, we focus solely on constraining the DM annihilation cross-section and coupling constants in the $ p $-wave scenario.

In the $ p $-wave scenario, the results of our analysis are summarized in Fig.~\ref{fig:fermi_con}. The most stringent constraints, derived from Fermi-LAT data near the GC, are as follows: For $ m_{h_s} = 0.3 \, \text{GeV} $ and $ m_{\chi} = 0.5 \, \text{GeV} $, the coupling constant $ c_s $ is constrained to $ 5.5 \times 10^{-5} $, while the predicted value from thermal relic abundance is $ 0.071 $. For $ m_{h_p} = 0.3 \, \text{GeV} $ and $ m_{\chi} = 0.5 \, \text{GeV} $, the coupling constant $ c_p $ is constrained to $ 8.9 \times 10^{-5} $, and the relevant value to fit thermal relic abundance is at $ 0.134 $. There is a close relationship between the magnitude of the flux and the annihilation cross-section $\sv$. Generally, when the flux is larger, the constraints on the coupling constants $c_{s,p}$ are also strengthened. Furthermore, an increase in the annihilation cross-section typically accompanies an enhancement of the flux. We find that selecting a mediator with a smaller mass can significantly increase the annihilation cross-section, thereby enhancing the signal's overall detection capability. In the scalar mediator case, the annihilation cross-section is larger than that in the pseudoscalar mediator case for the same parameter settings; thus, the constraints on $c_s$ are correspondingly more stringent than those on $c_p$.

In the near future, the Very Large
Gamma-ray Space Telescope (VLAST)~\cite{Fan:2022vlast,Pan:2024adp}, one of the few proposed next-generation space-based gamma-ray detectors, will have a detection area 5 times larger than Fermi-LAT and provide considerable help in
probing the sub-GeV DM. 
It is worth noting that the detection capabilities of the VLAST cover a broad energy range from below $1$ MeV to above $1$ TeV, which will greatly expand our detection range for DM masses and enable us to further investigate the forbidden annihilation scenario.

\section*{Acknowledgments}
We thank Mei-Wen Yang and Tian-Peng Tang for the helpful discussions. CTL and XYL are supported by the National Natural Science Foundation of China (NNSFC) under grant No.~12335005 and the Special funds for postdoctoral overseas recruitment, Ministry of Education of China. ZQX is supported by the National Key Research and Development Program of China (No. 2022YFF0503304).

\appendix{}
\section{DM self-interaction in minimal Higgs portal DM models}
\label{sec:self}

Considering Dirac-type DM, the DM self-interaction process involves the following three channels, $\chi\bar{\chi}\to\chi\bar{\chi}$, $\chi\chi\to\chi\chi$ and $\bar{\chi}\bar{\chi}\to\bar{\chi}\bar{\chi}$. When the mediator is a pseduoscalar, the cross-section of  $\chi\bar{\chi}\to\chi\bar{\chi}$ channel is,
\begin{equation}
\begin{aligned} \sigma(\chi\bar{\chi}\to\chi\bar{\chi})=\frac{(c_{p}/2)^4}{16\pi s^2 R_\chi^2}\left[\frac{sR_\chi^2(s R_\chi^2+2m_{h_p}^2)}{s R_\chi^2+m_{h_p}^2}+\frac{s^3 R_\chi^2}{(s-m_{h_p}^2)^2+m_{h_p}^2\Gamma_{h_p}^2}\right.\\
\left. +2m_{h_p}^2\ln(\frac{m_{h_p}^2}{s R_\chi^2+m_{h_p}^2})-
\frac{s m_{h_p}^2\ln(\frac{m_{h_p}^2}{s R_\chi^2+m_{h_p}^2})-s^2 R_\chi^2}{\sqrt{(s-m_{h_p}^2)^2+m_{h_p}^2\Gamma_{h_p}^2}}\right],
\end{aligned}
\end{equation}
for the channel $\chi\chi\to\chi\chi$ and $\bar{\chi}\bar{\chi}\to\bar{\chi}\bar{\chi}$, the cross-section is
\begin{equation}
\begin{aligned} \sigma(\chi\chi\to\chi\chi)=\sigma(\bar{\chi}\bar{\chi}\to\bar{\chi}\bar{\chi})=\frac{(c_{p}/2)^4(5m_{h_p}^2+3s R_\chi^2)(1-\frac{m_{h_p}^2}{sR_\chi^2+m_{h_p}^2}+\frac{2m_{h_p}^2\ln{\frac{m_{h_p}^2}{sR_\chi^2+m_{h_p}^2}}}{sR_\chi^2+2m_{h_p}^2})}{32\pi s^2 R_\chi^2}.
\end{aligned}
\end{equation}
Here $R_\chi = \sqrt{1-\frac{4m_\chi^2}{s}}$, $s\equiv E_{\rm cm}^2$ and $\Gamma_{h_{s,p}}$ is the decay width of mediator $h_{s,p}$.

When mediator is a scalar, cross-sections for the DM self-interaction process are
\begin{align}
&\sigma(\chi\chi\to\chi\chi)=\sigma(\bar{\chi}\bar{\chi}\to\bar{\chi}\bar{\chi})=\nonumber \\
&\frac{(c_{s}/2)^4}{32 \pi s (s - 4 m_{\chi}^2)} \left( -\frac{(4 m_{\chi}^2 - s) (5 m_{h_s}^4 + 32 m_{\chi}^4 + m_{h_s}^2 (-28 m_{\chi}^2 + 3 s))}{m_{h_s}^2 (m_{h_s}^2 - 4 m_{\chi}^2 + s)} \right. \nonumber \\
& \left. + \frac{2 (5 m_{h_s}^4 + 32 m_{\chi}^4 - 4 m_{\chi}^2 s + m_{h_s}^2 (-28 m_{\chi}^2 + 3 s)) \log \left( \frac{m_{h_s}^2}{m_{h_s}^2 - 4 m_{\chi}^2 + s} \right)}{2 m_{h_s}^2 - 4 m_{\chi}^2 + s} \right), 
\end{align}

\begin{align}
&\sigma(\chi\bar{\chi}\to\chi\bar{\chi})=\nonumber \\
& \frac{(c_{s}/2)^4}{16 \pi s (s - 4 m_{\chi}^2)}\left( -8 m_{\chi}^2 + \frac{16 m_{\chi}^4}{m_{h_s}^2} + s \left(2 - \frac{s}{m_{h_s}^2 - 4 m_{\chi}^2 + s}\right) \right. \nonumber \\
&\quad - \frac{(4 m_{\chi}^2 - s)^3}{(m_{h_s}^2 - s)^2 + m_{h_s}^2 \Gamma_{h_s}^2} + \frac{16 m_{\chi}^4}{\sqrt{(m_{h_s}^2 - s)^2 + m_{h_s}^2 \Gamma_{h_s}^2}}  - \frac{s^2}{\sqrt{(m_{h_s}^2 - s)^2 + m_{h_s}^2 \Gamma_{h_s}^2}} \nonumber \\
&\quad + \left. \frac{4 m_{\chi}^2 (4 m_{\chi}^2 - s - 2 \sqrt{(m_{h_s}^2 - s)^2 + m_{h_s}^2 \Gamma_{h_s}^2}) + m_{h_s}^2 (-4 m_{\chi}^2 - s + 2 \sqrt{(m_{h_s}^2 - s)^2 + m_{h_s}^2 \Gamma_{h_s}^2})}{\sqrt{(m_{h_s}^2 - s)^2 + m_{h_s}^2 \Gamma_{h_s}^2}} \right. \nonumber \\
&\quad \left. \times \log \left(\frac{m_{h_s}^2}{m_{h_s}^2 - 4 m_{\chi}^2 + s}\right) \right).
\end{align}

\begin{figure*}[ht]
\centering
\includegraphics[width=6.0cm]{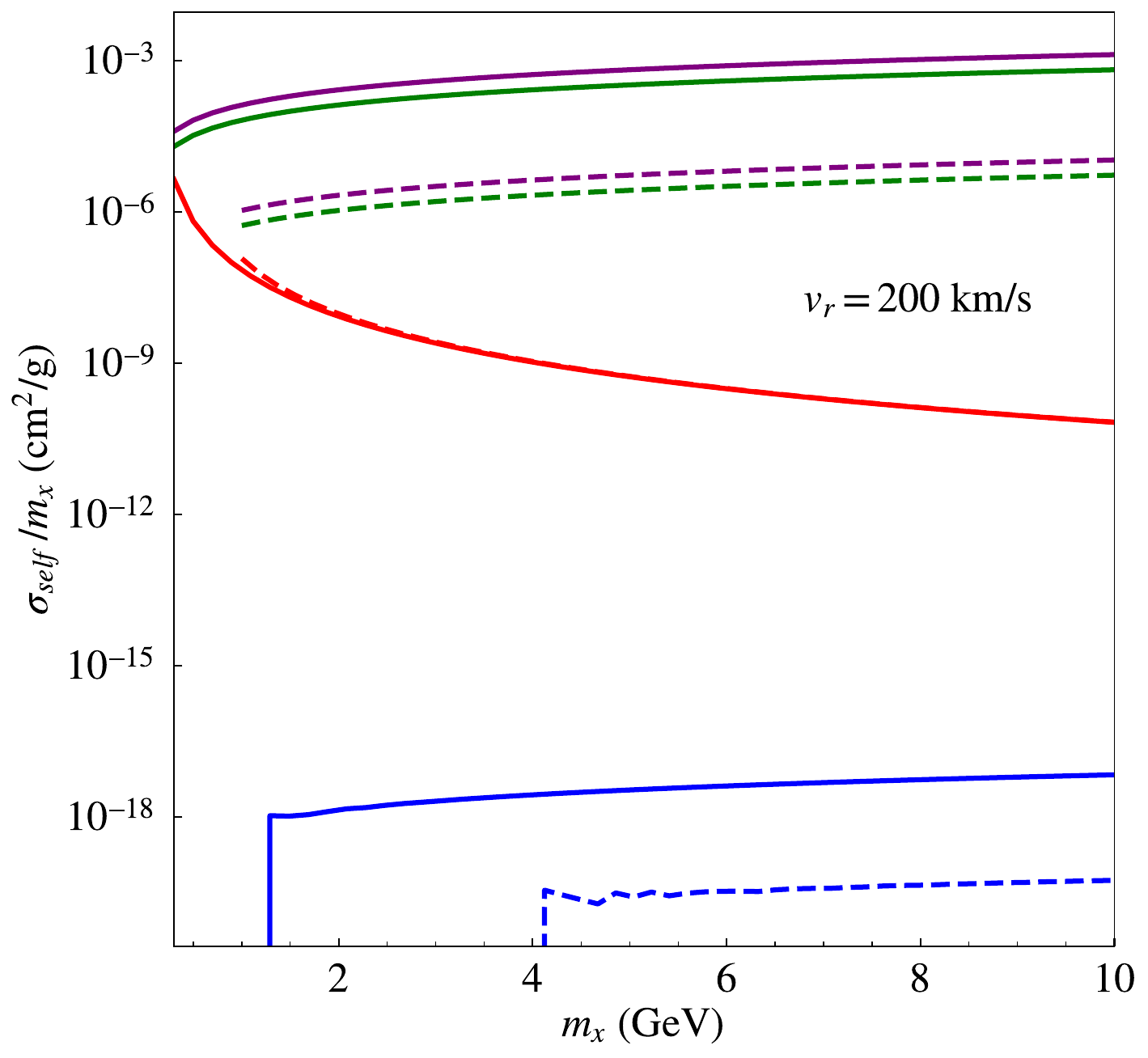}
\includegraphics[width=9.0cm]{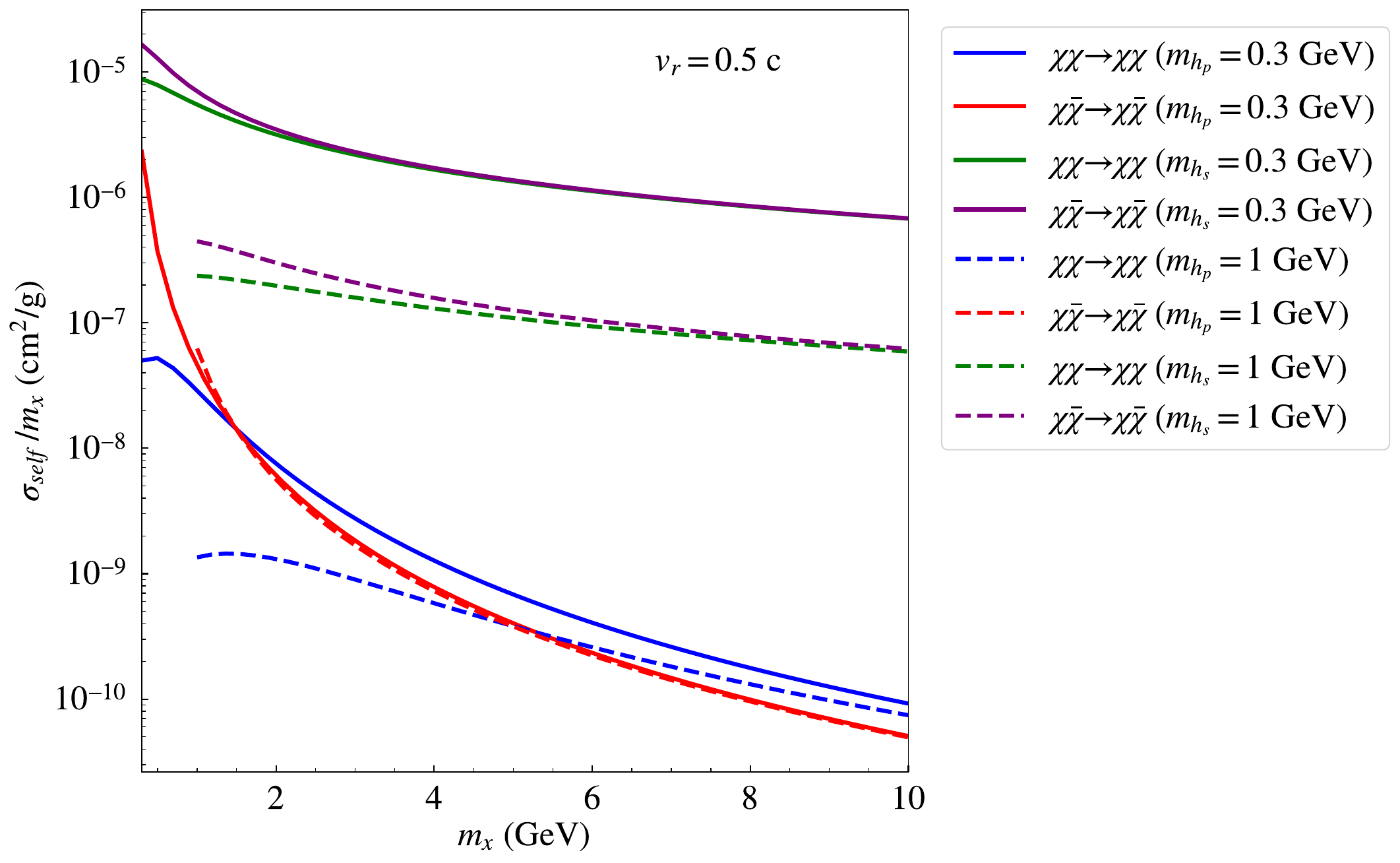}
 \caption{The cross-section of DM self-interaction in four possible channels with $v_{r}=200$ km/s (DM average velocity in the
Milky Way, left panel) and $v_{r}=0.5 c$ (the maximum speed after a SMBH acceleration, right panel). Solid lines represent the mediator mass $m_{h_s,h_p}=0.3 \gev$ and dashed lines represent the mediator mass $m_{h_s,h_p}=1 \gev$. }
 \label{fig:self_cs}
\end{figure*}

In Fig~\ref{fig:self_cs}, we present the variation of the DM self-interaction scattering cross-section for different channels as a function of DM mass under two scenarios: low velocities of DM particles (left panel) and velocities accelerated to half the speed of light (right panel). We find that, regardless of the scenario, the DM self-interaction scattering cross-section in minimal Higgs portal DM model is significantly smaller than the cross-section required to explain some small-scale structure problems for strong self-interactions $\sigma_{\text{self}}/m_{\chi} \approx 0.1 {\text{cm}}^2/{\text{g}}$~\cite{Shapiro:2014oha}. Therefore, we will not discuss the case of strong  self-interaction DM in this study.

\section{Electron, positron and neutrino spectra in the mediator rest frame}
\label{sec:eve}

\begin{figure}[tb]
    \includegraphics[width=0.45\textwidth]{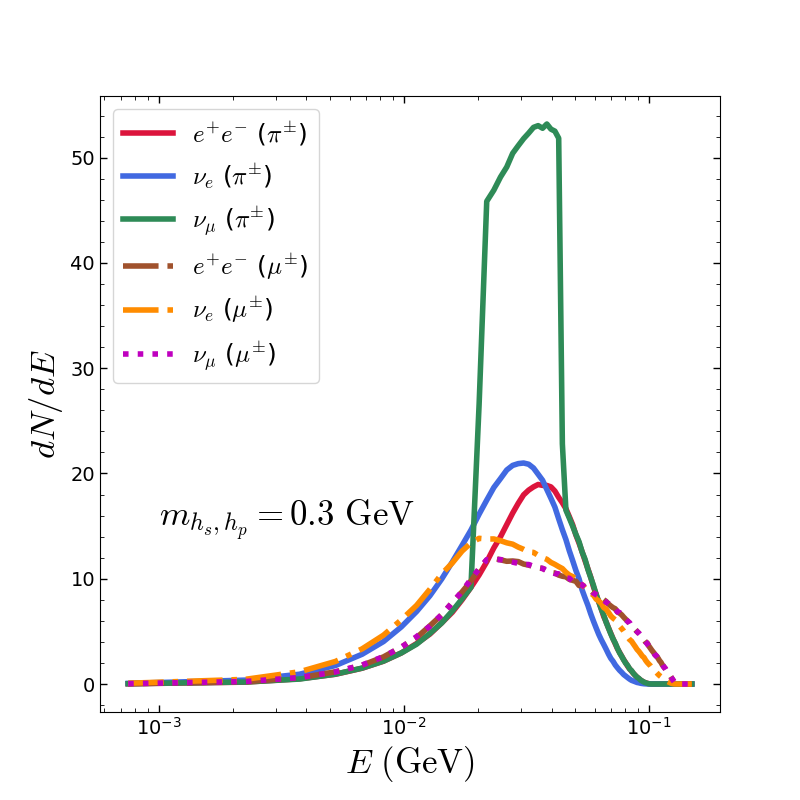}
    \includegraphics[width=0.45\textwidth]{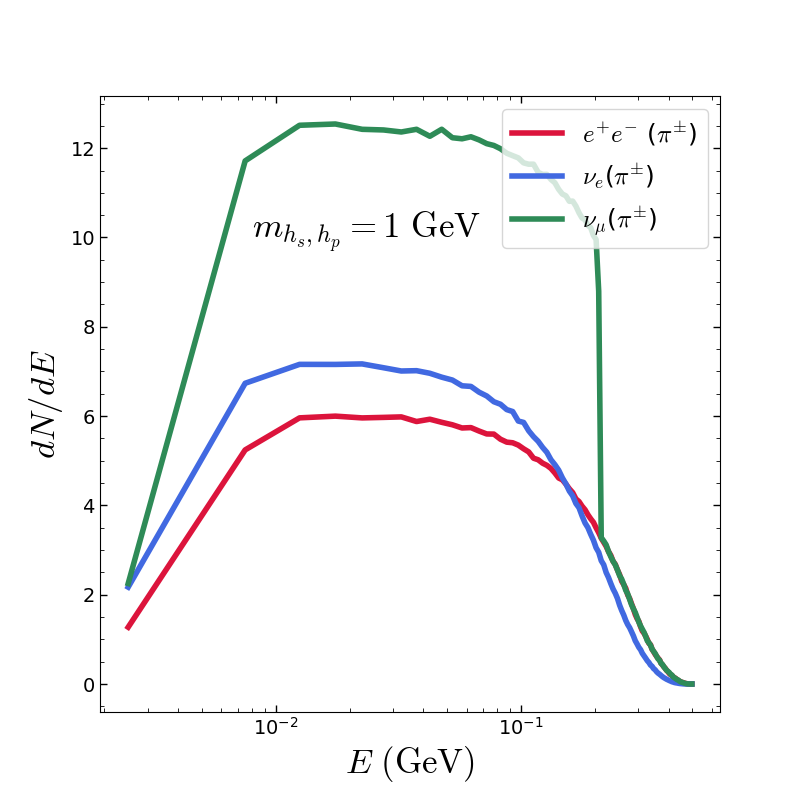}
	\caption{\label{fig:med}Electron, positron and neutrino spectra after multiplying by the decay branching ratios in the mediator rest frame. The left panel shows a fixed mediator mass of $0.3$ GeV, while the right panel shows a fixed mediator mass of $1$ GeV. The solid lines represent the generation of $\pi^{\pm}$ channels, and the dashed lines represent the generation of $\mu^{\pm}$ channels. }
\end{figure}

As shown in Fig~\ref{fig:med}, we present the energy spectra of electrons, positrons and neutrinos produced from the channels $ \pi^+ \pi^- $ (solid lines) and $ \mu^+ \mu^- $ (dashed lines) under two mediator mass scenarios: $ m_{h_s,h_p} = 0.3 \, \text{GeV} $ (left panel) and $ 1 \, \text{GeV} $ (right panel), after considering the decay branching ratios. The simulation method used here is identical to that employed for generating the photon energy spectrum. First, we utilized Pythia8 to generate the energy spectra of electrons, positrons, and neutrinos arising from various decay channels ($ h_{s,p}\to \pi^{+}\pi^{-} $ and $ h_{s,p} \to \mu^{+}\mu^{-} $) in the rest frame of the mediators. Subsequently, these spectra were weighted by their corresponding branching ratios to obtain the final results, as shown in Fig~\ref{fig:med}. We observe that, after incorporating the branching ratios, the energy spectrum from the $ \mu^+ \mu^- $ channel is suppressed, particularly at $ m_{h_s,h_p} = 1 \, \text{GeV} $, where the shape from the $ \mu^+ \mu^- $ channel closely resembles that of the $ \pi^+ \pi^- $ channel, contributing negligibly. Therefore, the $ \mu^+ \mu^- $ channel is not displayed in the right pane of Fig.~\ref{fig:med}. 

\bibliography{refs}
\end{document}